\documentclass[11pt,a4paper]{article}

\usepackage{geometry}
\usepackage{threeparttable}
\usepackage{booktabs}
\usepackage{amsfonts}
\usepackage{amsmath}
\usepackage[pdftex]{graphicx}
\usepackage{bm} 
\usepackage{mathrsfs}  
\usepackage{amssymb}
\usepackage{subcaption} 
\usepackage{algorithm}    
\usepackage{algorithmic}  

\newtheorem{assumption}{Assumption}
\newtheorem{lemma}{Lemma}
\newtheorem{theorem}{Theorem}

\usepackage{ragged2e}
\usepackage{tabularx}
\usepackage{caption}
\usepackage{xcolor}

\begin{document}

\title{Resilient Hierarchical Power Control for Hybrid GFL/GFM Microgrids Under Mixed Cyber-Attacks and Physical Constraints
\thanks{Lifu Ding was with State Key Laboratory of Industrial Control Technology, College of Control Science and Engineering, Zhejiang University, Hangzhou 310027, China. \\
Chunhui Hou was with College of Control Science and Engineering, Zhejiang University, Hangzhou 310027, China.\\
Qinmin Yang was with State Key Laboratory of Industrial Control Technology, College of Control Science and Engineering, Zhejiang University, Hangzhou 310027, China, Huzhou Industrial Control Technology Research Institute, Huzhou 313000, China, and Zhejiang Key Laboratory of Decision Intelligence, Hangzhou 310027, China.\\
Yutong Li was with the Research Institute of China Southern Power Grid Co., Ltd., Guangzhou 510663, China.}
\thanks{This work was financially supported by the China Southern Power Grid under Grant ZBKJXM20240023}
}

\author{Lifu Ding , Chunhui Hou, Yutong Li, and Qinmin Yang \thanks{Corresponding Author(Email: qmyang@zju.edu.cn)} 
}




%


\maketitle

    
\begin{abstract}
Hybrid microgrids integrating Grid-Following (GFL) and Grid-Forming (GFM) inverters present complex control challenges arising from the decoupling between long-term economic dispatch and real-time dynamic regulation, as well as the distinct physical limitations of heterogeneous inverters under cyber uncertainties. This paper proposes a Resilient Hierarchical Power Control (RHPC) strategy to unify these conflicting requirements within a cohesive framework. A standardized power increment mechanism is developed to bridge the tertiary and secondary layers, ensuring that real-time load fluctuations are compensated strictly according to the optimal economic ratios derived from the tertiary layer. To address the strict active power saturation constraints of GFL units, a dynamic activation scheme coupled with projection operators is introduced, which actively isolates saturated nodes from the consensus loop to prevent integrator wind-up and preserve the stability of the GFM backbone. Furthermore, the proposed framework incorporates a multi-scale attention mechanism and LSTM-based predictors into the secondary control protocol, endowing the system with robustness against unbounded False Data Injection (FDI) attacks and packet losses. Rigorous theoretical analysis confirms that the system achieves Uniformly Ultimately Bounded (UUB) convergence, and simulations on a modified IEEE 33-bus system demonstrate that the proposed strategy significantly improves power sharing accuracy and operational resilience in both grid-connected and islanded modes compared to conventional methods.
\end{abstract}


\vspace{-8pt}

\section{Introduction}
\label{sec:intro}

The global pursuit of carbon neutrality has accelerated the transformation of modern power systems towards high penetrations of renewable energy sources (RESs) \cite{liu2022challenges}. Microgrids (MGs), capable of operating in both grid-connected and islanded modes, have emerged as a fundamental building block for integrating distributed generators (DGs) and facilitating flexible demand response \cite{mariam2016microgrid}. While early microgrid designs predominantly relied on Grid-Following (GFL) inverters to track maximum power points, the diminishing system inertia in islanded modes has necessitated the adoption of Grid-Forming (GFM) inverters to provide essential voltage and frequency support \cite{GFIA, Duality}. Consequently, hybrid microgrids comprising heterogeneous GFL and GFM units are becoming the mainstream architecture, aiming to balance renewable utilization efficiency with system stability \cite{Hybrid_Review1, Hybrid_Review2}. Recent studies further highlight that the dynamic interaction between GFL and GFM units can significantly impact system damping and transient stability, necessitating advanced coordinated control strategies \cite{rathnayake2021grid, lin2020research}.

To manage the complexity of such clusters, a hierarchical control framework is widely standardized, consisting of primary, secondary, and tertiary layers \cite{hennane2022primary}. The tertiary layer typically performs Optimal Power Flow (OPF) to determine economic dispatch setpoints based on long-term forecasts \cite{DPTCDC, Ref_Chapter3}, while the secondary layer compensates for real-time voltage/frequency deviations and shares power among DGs via distributed consensus protocols \cite{lin2015reach}. 
However, a critical coordination gap persists between these two upper layers in existing literature. Conventional secondary control strategies often employ fixed droop coefficients or capacity-based ratios for power sharing \cite{CCfGFLGFM}. This rigid sharing mechanism can severely conflict with the optimal dispatch commands issued by the tertiary layer. For instance, if the tertiary layer schedules an Energy Storage System (ESS) to charge (negative power) for arbitrage, a traditional secondary controller might force it to discharge during a sudden load increase, thereby violating the economic strategy and state-of-charge (SoC) management \cite{TSCLEO, Two_layer}. Therefore, designing a dynamic compensation mechanism that strictly adheres to the economic proportions defined by the tertiary layer is a pressing necessity.

Beyond hierarchical coordination, the physical heterogeneity of GFL and GFM units introduces distinct control challenges, particularly concerning physical constraints. GFM units, acting as voltage sources, generally possess higher overload capabilities, whereas GFL units function as current sources with strict active power saturation limits \cite{aljarrah2024issues}. Most existing distributed control protocols assume DGs operate within linear regions \cite{OOGCBOC, CBDCS}. In practice, when a GFL unit reaches its output limit, the discrepancy between the computed virtual control signal and the saturated physical output leads to the "integrator wind-up" phenomenon. This not only degrades the dynamic performance of power sharing but can also destabilize the entire communication network by propagating erroneous state information \cite{TRENG}. Although some works have explored constrained consensus problems using projection operators \cite{nedic2010constrained} or barrier functions \cite{kundu2019distributed}, few have addressed the specific scenario of hybrid GFL/GFM interaction where GFLs must be actively isolated from the sharing loop upon saturation while GFM units maintain system balance.

Furthermore, the dependence of distributed secondary control on communication networks exposes hybrid microgrids to severe cyber-physical threats. The Cyber-Physical System (CPS) nature of MGs makes them vulnerable to various anomalies, including False Data Injection (FDI) attacks, Denial of Service (DoS) attacks (manifesting as Packet Loss, PL), and time-varying delays \cite{DMCASCS, DRSCDC}. Malicious FDI attacks can manipulate consensus variables to mislead power dispatch \cite{RACR}, while PL and delays disrupt the synchronization required for stable operation. While robust control strategies against specific attacks have been developed—such as trust-based detection \cite{DEBRSC}, observer-based reconstruction \cite{FDIARD}, and mean-subsequence-reduced (MSR) algorithms \cite{RCFDTMS}—these methods often rely on restrictive assumptions (e.g., bounded attack signals or homogeneous agent dynamics). More sophisticated stealthy attacks, which utilize system knowledge to bypass conventional detectors, pose an even greater risk to distributed coordination \cite{zuo2020distributed}. A comprehensive framework capable of tolerating unbounded FDI and mixed communication faults within a heterogeneous GFL/GFM network remains largely unexplored.

Motivated by the "trilemma" of hierarchical conflict, physical constraints, and cyber-fragility, this paper proposes a Resilient Hierarchical Power Control (RHPC) strategy for hybrid GFL/GFM microgrids. Integrating a data-driven tertiary optimization layer with a robust secondary control loop, this paper makes the following contributions:

\begin{enumerate}
    \item \textbf{Standardized Power Increment for Hierarchical Coordination:} A standardized power increment is introduced as the consensus variable. Unlike traditional absolute power sharing, this variable represents the normalized deviation relative to the tertiary setpoints. This mechanism ensures that real-time load fluctuations are shared strictly according to the economic ratios optimized by the tertiary layer, effectively bridging the timescale gap between economic dispatch and dynamic regulation.
    
    \item \textbf{Active Constraint Handling via Activation Function:} To address the distinct physical limits of GFL units, a dynamic \textit{Activation Function} combined with a \textit{Projection Operator} is devised. This mechanism serves as a logical switch: when a GFL unit saturates, its activation function zeroes out, effectively removing it from the consensus topology. This prevents wind-up issues and ensures that the remaining GFM units automatically compensate for the power deficit, maintaining system stability.
    
    \item \textbf{Resilience Against Mixed Cyber-Attacks:} A distributed robust controller is developed to tolerate mixed faults, including unbounded FDI attacks, packet losses, and delays. By incorporating a multi-scale attention mechanism for precise anomaly isolation and an LSTM-based predictor for data compensation, the proposed strategy guarantees Uniformly Ultimately Bounded (UUB) stability for the hybrid system.
\end{enumerate}

The remainder of this paper is organized as follows. Section \ref{sec:problem} formulates the problem regarding load fluctuations and constraints in hybrid MGs. Section \ref{sec:control} details the proposed hierarchical strategy and the design of the robust controller. Stability analysis is provided in Section \ref{sec:stability}. Simulation results on an IEEE 33-bus system are presented in Section \ref{sec:sim}, and Section \ref{sec:concl} concludes the paper.

\vspace{-2pt}
\section{Problem Formulation}
\label{sec:problem}

This section establishes the comprehensive modeling framework for the hybrid microgrid. We first delineate the heterogeneous physical characteristics of GFL and GFM inverters, which underpin the distinct constraint formulations. Subsequently, the hierarchical power fluctuation model is derived to bridge the timescale gap between economic dispatch and dynamic regulation. Finally, a rigorous cyber-layer fault model is presented to capture the mixed uncertainties in the communication network.

Consider a hybrid microgrid comprising $N$ distributed generators (DGs). The set of DGs is partitioned into GFL-based units, denoted by $\mathcal{B}_{dl}$, and GFM-based units, denoted by $\mathcal{B}_{dm}$, satisfying $\mathcal{B}_{dl} \cup \mathcal{B}_{dm} = \mathcal{B}_d$ and $\mathcal{B}_{dl} \cap \mathcal{B}_{dm} = \emptyset$. The Point of Common Coupling (PCC), which manages the power exchange with the main grid, is indexed as node $n_d+1$.

\vspace{-3pt}
\subsection{Physical Characteristics and Heterogeneity}
The control challenges in hybrid microgrids stem fundamentally from the distinct physical behaviors of the inverters.
GFL inverters are typically controlled as current sources, synchronizing to the grid via Phase-Locked Loops (PLLs). They are designed to strictly track active and reactive power setpoints but lack inherent inertia. Crucially, their power electronic components impose strict thermal limits, resulting in a bounded feasible operating region $[\underline{P}_i, \overline{P}_i] \times [\underline{Q}_i, \overline{Q}_i]$. Exceeding these limits can trigger overcurrent protection and disconnect the unit.
In contrast, GFM inverters emulate voltage sources (e.g., via Virtual Synchronous Machine control), providing the voltage and frequency reference for the islanded grid. They possess a degree of short-term overload capability, acting as the slack bus to balance instantaneous power mismatches.
This physical heterogeneity necessitates a control strategy that explicitly respects the hard constraints of GFL units while leveraging the flexibility of GFM units.

\vspace{-3pt}
\subsection{Hierarchical Power Fluctuation Modeling}
The system operates under a hierarchical framework. At the tertiary layer (timescale $t_l$), an optimization algorithm (e.g., COSAC \cite{Ref_Chapter3}) solves a multi-objective Optimal Power Flow (OPF) problem to determine the base economic setpoints $P_{DG,i}$ and $Q_{DG,i}$. These setpoints minimize generation costs and network losses, potentially scheduling ESS units to charge (negative power) during low-price periods.

However, within the faster secondary control interval $t_s$ (indexed by $k$), the actual load demand deviates from the forecast. The total instantaneous load $P_L(k)$ and $Q_L(k)$ is decomposed as:
\begin{equation} \label{eq:load_decomp}
\begin{cases}
P_L(k) = P_{load}^s + P_{loss} + \Delta P_L(k), \\
Q_L(k) = Q_{load}^s + Q_{loss} + \Delta Q_L(k),
\end{cases}
\end{equation}
where $P_{load}^s, Q_{load}^s$ are the forecasted base loads, and $P_{loss}, Q_{loss}$ represent the quasi-steady-state losses. The terms $\Delta P_L(k)$ and $\Delta Q_L(k)$ capture the real-time stochastic fluctuations that must be compensated by the secondary control.

To define the compensation mechanism, the actual output of DG $i$ is formulated as:
\begin{equation} \label{eq:power_comp}
\begin{cases}
P_i(k) = P_{DG,i} + \Delta P_i(k), \\
Q_i(k) = Q_{DG,i} + \Delta Q_i(k).
\end{cases}
\end{equation}
Here, $\Delta P_i(k)$ and $\Delta Q_i(k)$ are the regulation variables. The global power balance in the secondary timescale requires:
\begin{equation} \label{eq:balance}
\sum_{i \in \mathcal{B}_d} \Delta P_i(k) + \Delta P_{pcc}(k) = \Delta P_L(k),
\end{equation}
where $\Delta P_{pcc}(k)$ denotes the deviation of PCC exchange power. In islanded mode, $P_{pcc}(k) \equiv 0$ implies $\Delta P_{pcc}(k) \equiv 0$.

For the stability of the control design, we impose a mild assumption on the load dynamics:
\begin{assumption} \label{asm:bounded_rate}
The rate of change of load fluctuations is bounded, i.e., $|\Delta P_L(k+1) - \Delta P_L(k)| \le \overline{\Delta P_L}$ and $|\Delta Q_L(k+1) - \Delta Q_L(k)| \le \overline{\Delta Q_L}$, where $\overline{\Delta P_L}, \overline{\Delta Q_L} > 0$ are known constants.
\end{assumption}

\vspace{-3pt}
\subsection{Standardized Increment as Consensus Variable}
A critical flaw in traditional secondary control approaches (e.g., capacity-based ratio sharing) is the potential conflict with tertiary economic dispatch. For instance, if the tertiary layer schedules an ESS to charge ($P_{DG,i} < 0$) while another generator discharges ($P_{DG,j} > 0$), a simple proportional sharing rule based on capacities might force the ESS to discharge during a load increase, violating the SoC management strategy.

To resolve this, we introduce the \textit{Standardized Power Increment} as the state variable $\bm{x}_i(k) = [x_{i,1}(k), x_{i,2}(k)]^T$:
\begin{equation} \label{eq:state_def}
x_{i,1}(k) = \frac{\Delta P_i(k)}{|P_{DG,i}|}, \quad x_{i,2}(k) = \frac{\Delta Q_i(k)}{|Q_{DG,i}|}, \quad i \in \mathcal{B}_d.
\end{equation}
This definition implies that the real-time compensation $\Delta P_i(k)$ is proportional to the magnitude of the economic setpoint $|P_{DG,i}|$. Consequently, DGs with larger scheduled outputs (or charging rates) contribute more to regulation, preserving the relative economic ratios determined by the tertiary layer. The control objective transforms into a consensus problem:
\begin{equation} \label{eq:consensus_obj}
\lim_{k \to \infty} |x_{i,1}(k) - x_{j,1}(k)| \le \epsilon_p, \quad \forall i,j \in \mathcal{B}_d.
\end{equation}

Considering the strict limits of GFL units discussed in Section \ref{sec:problem}.A, the feasible state space for GFL units is constrained to a closed convex set $\mathcal{X}_i$:
\begin{equation} \label{eq:constraints}
\mathcal{X}_i = \left\{ \bm{x}_i \in \mathbb{R}^2 \ \bigg| \ \frac{\underline{P}_i - P_{DG,i}}{|P_{DG,i}|} \le x_{i,1} \le \frac{\overline{P}_i - P_{DG,i}}{|P_{DG,i}|} \right\}.
\end{equation}
For GFM units, owing to their role as the grid backbone, we assume an unbounded feasible region $\mathcal{X}_i = \mathbb{R}^2$ within the secondary control timescale. The dynamics of the standardized increment states are modeled as integrators:
\begin{equation} \label{eq:dynamics}
\bm{x}_i(k+1) = \bm{u}_i(k), \quad i \in \mathcal{B}_d,
\end{equation}
where $\bm{u}_i(k)$ is the control input.

\subsection{Mixed Cyber-Layer Fault Modeling}
The distributed realization of \eqref{eq:consensus_obj} relies on a communication network $\mathcal{G}(\mathcal{V}, \mathcal{E})$. We consider a realistic mixed fault model encompassing False Data Injection (FDI), Packet Loss (PL), and transmission delays.

\subsubsection{Unbounded False Data Injection}
FDI attacks are modeled as malicious signals superimposed on the communication links. If link $(j, i)$ is compromised, the state received by node $i$ is:
\begin{equation} \label{eq:fdi_model}
\tilde{\bm{x}}_{ij}(k) = \bm{x}_j(k-\tau_{ij}) + \beta_{ij}(k) \bm{\phi}_{ij}^m(k),
\end{equation}
where $\tau_{ij}$ is the delay, $\beta_{ij}(k) \in \{0, 1\}$ is the attack indicator, and $\bm{\phi}_{ij}^m(k)$ is the injected data.
Distinct from standard bounded noise, we consider \textit{unbounded} and \textit{stealthy} FDI attacks. Specifically, the injected signal $\bm{\phi}_{ij}^m(k)$ may have arbitrary magnitude but is often designed to mimic normal system dynamics (e.g., low-frequency trends) to evade conventional threshold-based detectors. Following the $f$-local attack model \cite{RACR}, we assume the number of compromised incoming links for any node is at most $f$.

\subsubsection{Probabilistic Packet Loss and Delay}
Network congestion or DoS attacks can lead to packet dropouts. This is modeled by a Bernoulli process, where the probability of successful transmission is $1-\alpha$, with $\alpha \in [0, 1)$ being the upper bound of the loss probability. Additionally, bounded time-varying delays $\tau_{ij}(k) \le \bar{\tau}$ are inherent in the signal transmission and processing.

The comprehensive problem is thus to design a distributed protocol $\bm{u}_i(k)$ that achieves the consensus objective \eqref{eq:consensus_obj} and maintains stability (UUB) despite the presence of constrained feasible sets $\mathcal{X}_i$ for GFL units and mixed cyber-faults in the communication layer.

\vspace{-8pt}
\section{Hierarchical Coordinated Control Strategy}
\label{sec:control}

To address the challenges of conflicting control layers, strict physical constraints, and cyber uncertainties, a robust hierarchical control strategy is proposed. This framework synthesizes a long-term economic dispatch optimizer with a resilient, constraint-aware secondary control loop. By employing a standardized increment interface and a hybrid continuous-discrete saturation management scheme, the strategy ensures seamless coordination across timescales while guaranteeing system stability under adverse conditions. The overall control logic is summarized in Algorithm \ref{alg:control_logic}.

\vspace{-8pt}
\subsection{Tertiary-Secondary Information Exchange Framework}
\begin{figure*}[htbp]
    \centering
    \includegraphics[width=0.7\linewidth]{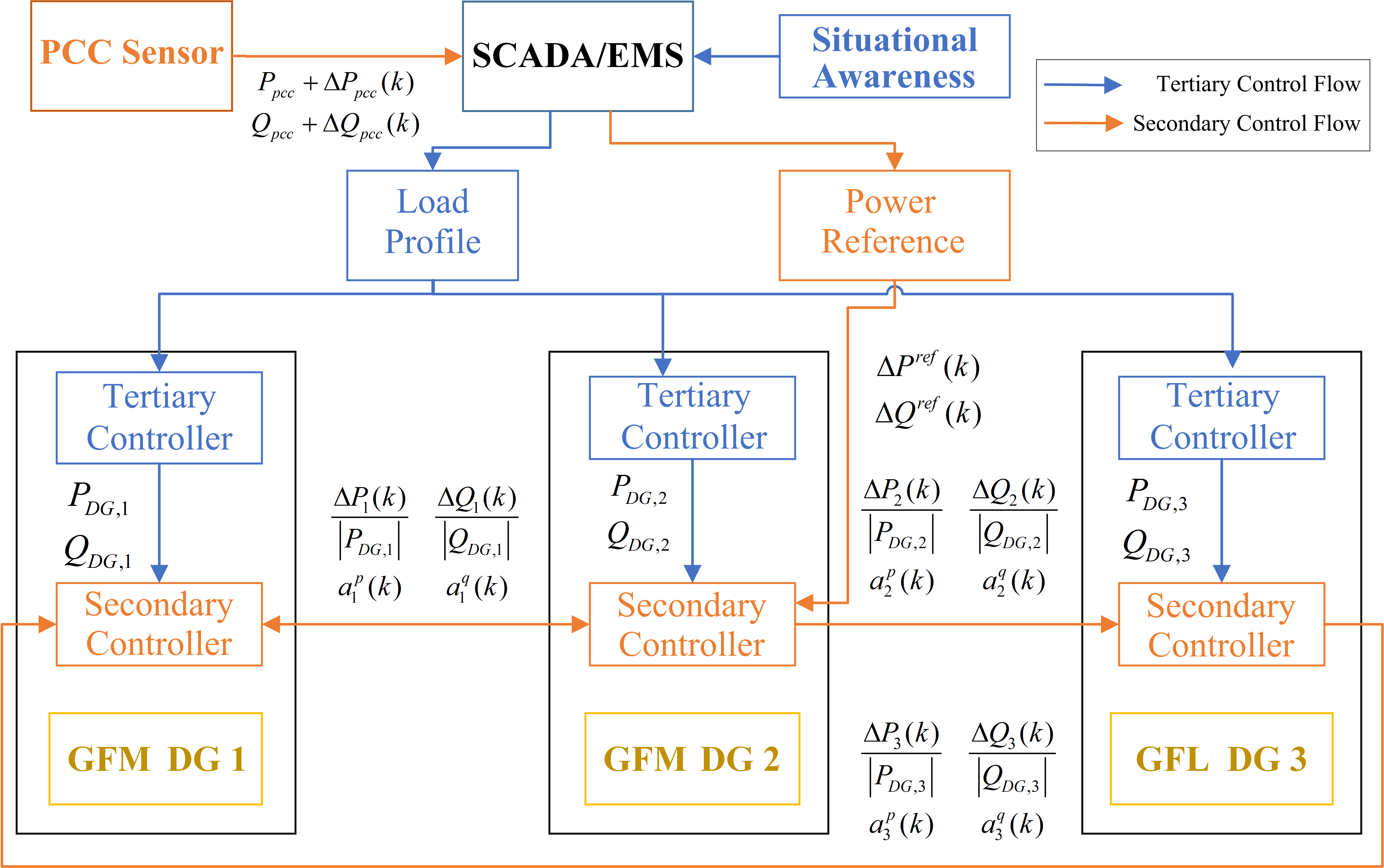}
    \caption{The proposed tertiary-secondary hierarchical coordinated control framework.}
    \label{fig:control_arch}
\end{figure*}

The proposed coordination relies on a hierarchical communication structure, as illustrated in Fig. \ref{fig:control_arch}. 
The SCADA/EMS functions as a centralized coordinator (indexed as node 0) for the tertiary layer but operates as a ``virtual leader'' for the distributed secondary layer.

At the tertiary timescale $t_l$, the SCADA/EMS executes an optimal power flow algorithm (e.g., COSAC \cite{Ref_Chapter3}) to compute the economic base setpoints $P_{DG,i}$ and $Q_{DG,i}$. These setpoints reflect the optimal trade-off between generation costs, network losses, and ESS degradation. At the beginning of each interval, these values are broadcast to the local controllers of all DGs to serve as the normalization baseline for the standardized increment defined in \eqref{eq:state_def}.

During the interval, the SCADA/EMS continuously monitors the power exchange at the Point of Common Coupling (PCC). It computes a dynamic reference trajectory for the standardized increment, $\Delta \bm{x}^{ref}(k)$, to guide the microgrid in compensating for the total real-time load mismatch $\Delta P_L$. The update rule is defined as:
\begin{equation} \label{eq:ref_update}
\Delta \bm{x}^{ref}(k+1) = \frac{1}{|\mathcal{N}_0|+1} \left[ \sum_{i \in \mathcal{N}_0} \bm{x}_i(k) + \bm{A}_{pcc}^{ac}(k) \bm{x}_{pcc}(k) \right],
\end{equation}
where $\mathcal{N}_0 \subseteq \mathcal{B}_{dm}$ denotes the set of pinning nodes (GFM units) communicating directly with the SCADA/EMS, and $\bm{x}_{pcc}(k)$ represents the standardized increment of the PCC power. The term $\bm{A}_{pcc}^{ac}(k)$ is a binary activation parameter indicating the availability of reliable PCC measurements. This dynamic reference generation ensures that the secondary control layer acts as a "follower" to the tertiary economic intent, scaling the dispatch proportionally rather than deviating from it.

To ensure the solvability of the consensus problem, the communication topology $\mathcal{G}$ is assumed to satisfy the following connectivity condition:
\begin{assumption} \label{asm:topology}
The communication subgraph formed by the SCADA/EMS (node 0) and the GFM nodes ($\mathcal{B}_{dm}$) contains a spanning tree rooted at node 0. Furthermore, for every GFL node $i \in \mathcal{B}_{dl}$, there exists a directed path from at least one node in $\mathcal{B}_{dm}$ to $i$.
\end{assumption}

\vspace{-8pt}
\subsection{Resilient Control Protocol with Multi-Scale Attention}
Discrete-time distributed control protocols are designed to handle the distinct physical characteristics of GFL/GFM inverters while tolerating mixed cyber threats. The design integrates a resilient weighting mechanism directly into the consensus update.

For \textbf{GFL-based DGs} ($i \in \mathcal{B}_{dl}$), which function as current sources with strict active power saturation limits, the control input utilizes a projection-based consensus mechanism:
\begin{equation} \label{eq:ctrl_gfl}
\bm{u}_i(k) = \mathscr{P}_{\mathcal{X}_i} \left[ (1-c)\bm{x}_i(k) + c \sum_{j \in \mathcal{N}_i} a_{ij}(k) \bm{\tilde{x}}_{ij}(k-\tau_{ij}) \right],
\end{equation}
where $c$ is the step size. $\mathscr{P}_{\mathcal{X}_i}(\bm{z}) = \arg \min_{\bm{y} \in \mathcal{X}_i} \|\bm{z} - \bm{y}\|$ is the projection operator onto the closed convex set $\mathcal{X}_i$. This operator rigorously enforces the physical safety bounds $[\underline{P}_i, \overline{P}_i]$ defined in \eqref{eq:constraints}, ensuring that the virtual state $\bm{x}_i$ never demands a physical output beyond the inverter's capability.

For \textbf{GFM-based DGs} ($i \in \mathcal{B}_{dm}$), which act as the voltage backbone with higher overload capabilities, the protocol incorporates a reference tracking term and an activation matrix to handle the potential saturation of neighbors:
\begin{align} \label{eq:ctrl_gfm}
\bm{u}_i(k) = & (1-c)\bm{x}_i(k) + c \sum_{j \in \mathcal{N}_i} \bm{A}_j^{ac}(k) a_{ij}(k) \bm{\tilde{x}}_{ij}(k-\tau_{ij}) \nonumber \\
& + c b_i (\Delta \bm{x}^{ref}(k) - \bm{x}_i(k)),
\end{align}
where $b_i > 0$ if DG $i \in \mathcal{N}_0$, and $0$ otherwise.

The term $a_{ij}(k)$ represents the \textit{Resilient Weight}. To effectively isolate unbounded FDI attacks, specifically those using stealthy low-frequency signals, a \textbf{Multi-scale Attention Mechanism} is employed. The state discrepancy is evaluated across three temporal scales: instantaneous value, moving average (short-term trend), and historical accumulation (long-term memory). The feature vector $r_{ij}^{(s)}(k)$ for scale $s \in \{1, 2, 3\}$ is defined as:
\begin{equation}
r_{ij}^{(s)}(k) = \begin{cases}
\bm{\tilde{x}}_{ij}(k), & s=1 \\
\frac{1}{W}\sum_{m=1}^{W} \bm{\tilde{x}}_{ij}(k-m), & s=2 \\
(1-\gamma)r_{ij}^{(3)}(k-1) + \gamma \bm{\tilde{x}}_{ij}(k), & s=3
\end{cases}
\end{equation}
where $W$ is the window size and $\gamma$ is a smoothing factor. The attention weight is then dynamically computed using a Softmax function over the aggregated feature distances:
\begin{equation} \label{eq:attention_weight}
a_{ij}(k) = \frac{\exp(-\sum_{s=1}^3 \sigma_s \|r_{ij}^{(s)}(k) - r_{i}^{(s)}(k)\|)}{\sum_{l \in \mathcal{N}_i} \exp(-\sum_{s=1}^3 \sigma_s \|r_{il}^{(s)}(k) - r_{i}^{(s)}(k)\|)},
\end{equation}
where $\sigma_s$ denotes the sensitivity parameter for each scale. This ensures that attacks deviating in any temporal characteristic are suppressed.

To mitigate Packet Loss (PL), an LSTM-based predictor is integrated. When a packet from neighbor $j$ is lost, the predictor generates an estimated state $\hat{\bm{x}}_{j}(k)$ using an input sequence $\mathcal{I}_j(k) = \{(\bm{\tilde{x}}_{ij}(k-m), t_{local}(k-m))\}_{m=1}^H$, where $H$ is the history horizon. This allows the controller to maintain operation even under intermittent communication failures.

\vspace{-8pt}
\subsection{Active Constraint Handling Mechanism}
A key innovation of this strategy is the \textbf{Activation Matrix} $\bm{A}_j^{ac}(k)$, which effectively creates a hybrid dynamical system where continuous control laws interact with discrete logic states.
When a GFL unit $j$ reaches its power limit (i.e., its state $\bm{x}_j$ hits the boundary of $\mathcal{X}_j$), it loses the capability to participate in proportional power sharing for \textit{additional} load fluctuations. Including such a "saturated" node in the consensus loop would introduce a persistent non-zero error, causing the "integrator wind-up" effect and dragging down the convergence of the entire network.

To prevent this, the activation matrix $\bm{A}_j^{ac}(k) = \text{diag}(a_{j,1}^{ac}, a_{j,2}^{ac})$ is defined as a state-dependent switching signal:
\begin{equation} \label{eq:activation}
a_{j,m}^{ac}(k) = 
\begin{cases} 
0, & \text{if } x_{j,m}(k) \in \partial \mathcal{X}_j \text{ (Boundary)}, \\
1, & \text{otherwise},
\end{cases}
\end{equation}
for $m \in \{1, 2\}$ corresponding to active and reactive power dimensions.

By multiplying the neighbor's state with $\bm{A}_j^{ac}(k)$ in Eq. \eqref{eq:ctrl_gfm}, the GFM units logically "disconnect" from the saturated GFL units in the consensus update graph. The saturated GFL unit becomes a boundary input node (clamped by the Projection Operator), while the remaining unsaturated GFM units automatically re-adjust their outputs to compensate for the remaining load mismatch $\Delta P_L$. This mechanism effectively transforms the non-linear physical saturation problem into a switching topology consensus problem, preserving the linearity and convergence properties of the active sub-network.

\begin{algorithm}[htbp]
\caption{Resilient Hierarchical Power Control Algorithm}
\label{alg:control_logic}
\begin{algorithmic}[1]
\REQUIRE Tertiary setpoints $P_{DG,i}$, limits $\mathcal{X}_i$.
\STATE \textbf{Initialization:} Set initial states $\bm{x}_i(0) = \mathbf{0}$.
\FOR{each time step $k$}
    \STATE \textbf{Step 1: Data Acquisition \& Pre-processing}
    \FOR{each neighbor $j \in \mathcal{N}_i$}
        \IF{Packet Received}
            \STATE Compute multi-scale features $r_{ij}^{(s)}(k)$.
            \STATE Calculate attention weights $a_{ij}(k)$ via Eq. \eqref{eq:attention_weight}.
        \ELSE
            \STATE Estimate state $\hat{\bm{x}}_{j}(k)$ using LSTM predictor.
            \STATE Use previous weights $a_{ij}(k) = a_{ij}(k-1)$.
        \ENDIF
    \ENDFOR
    
    \STATE \textbf{Step 2: Reference Update (if connected to SCADA)}
    \IF{$i \in \mathcal{N}_0$}
        \STATE Update $\Delta \bm{x}^{ref}(k)$ via Eq. \eqref{eq:ref_update}.
    \ENDIF
    
    \STATE \textbf{Step 3: Control Update Calculation}
    \IF{DG $i$ is GFL ($i \in \mathcal{B}_{dl}$)}
        \STATE Calculate $\bm{u}_i(k)$ via Eq. \eqref{eq:ctrl_gfl} with Projection $\mathscr{P}_{\mathcal{X}_i}$.
        \STATE Update Activation $\bm{A}_i^{ac}(k)$ based on boundary status.
    \ELSE
        \STATE Calculate $\bm{u}_i(k)$ via Eq. \eqref{eq:ctrl_gfm} using neighbors' $\bm{A}_j^{ac}(k)$.
    \ENDIF
    
    \STATE \textbf{Step 4: State \& Power Output Update}
    \STATE $\bm{x}_i(k+1) = \bm{u}_i(k)$.
    \STATE $P_i(k+1) = P_{DG,i} + \bm{x}_{i,1}(k+1)|P_{DG,i}|$.
\ENDFOR
\end{algorithmic}
\end{algorithm}

\section{Stability Analysis}
\label{sec:stability}

/

We define the global state vector $\bm{X}(k) = [\bm{x}_1(k)^T, \dots, \bm{x}_N(k)^T]^T$ and the synchronous reference vector $\bm{X}^{ref}(k) = \mathbf{1}_N \otimes \Delta \bm{x}^{ref}(k)$. The global synchronization error is formulated as $\bm{E}(k) = \bm{X}(k) - \bm{X}^{ref}(k)$.

\subsection{Preliminaries and Lemmas}

First, we invoke the fundamental non-expansive property of the projection operator, which ensures that the physical constraints on GFL units do not amplify the error energy.
\begin{lemma} \label{lem:projection}
Let $\mathcal{C} \subseteq \mathbb{R}^n$ be a non-empty closed convex set. The projection operator $\mathscr{P}_{\mathcal{C}}(\cdot)$ is non-expansive, satisfying the inequality:
\begin{equation}
\|\mathscr{P}_{\mathcal{C}}(\bm{x}) - \mathscr{P}_{\mathcal{C}}(\bm{y})\| \le \|\bm{x} - \bm{y}\|, \quad \forall \bm{x}, \bm{y} \in \mathbb{R}^n.
\end{equation}
\end{lemma}
\noindent \textit{Proof:} See \cite{Bertsekas_Convex}. \hfill $\blacksquare$

Next, we characterize the residual effect of cyber-attacks processed by the resilient weighting mechanism. Based on the multi-scale attention logic \eqref{eq:attention_weight}, large deviations caused by unbounded FDI attacks are suppressed, leaving only a bounded residual.
\begin{lemma} \label{lem:attack_bound}
Under the proposed multi-scale attention mechanism, the aggregate attack term $\bm{\Phi}_{att}(k)$ entering the control loop is bounded, i.e., $\|\bm{\Phi}_{att}(k)\| \le \xi_{att}$, where $\xi_{att}$ is a constant scalar dependent on the sensitivity parameters $\sigma_s$ and the maximum number of compromised links $f$.
\end{lemma}

\subsection{Main Stability Theorem}

\begin{theorem} \label{thm:stability_rigorous}
Consider the hybrid microgrid system governed by the dynamics \eqref{eq:power_comp}-\eqref{eq:dynamics}. Suppose Assumption \ref{asm:bounded_rate} (Bounded Load Rate) and Assumption \ref{asm:topology} (Network Connectivity) hold. Under the proposed hierarchical control protocols \eqref{eq:ctrl_gfl}-\eqref{eq:ctrl_gfm}, the synchronization error $\bm{E}(k)$ is Uniformly Ultimately Bounded (UUB). Specifically, there exists a finite time $K^*$ such that for all $k > K^*$, the error satisfies $\|\bm{E}(k)\| \le \mu$, where the bound $\mu$ is a function of the load fluctuation rate $\overline{\Delta P_L}$ and the attack residual $\xi_{att}$.
\end{theorem}

\noindent \textit{Proof:}

The proof is constructed by analyzing the evolution of the global error dynamics under switching topologies. Due to the state-dependent activation matrix $\bm{A}^{ac}(k)$, the communication topology switches among a finite set of graphs $\{\mathcal{G}_1, \dots, \mathcal{G}_M\}$. Let $\sigma(k) \in \{1, \dots, M\}$ denote the index of the active topology at time $k$.

\subsubsection{Local Error Evolution}
For GFL nodes $i \in \mathcal{B}_{dl}$, the control input involves the projection operator. Let $\bm{v}_i(k)$ denote the pre-projection update value derived from the consensus protocol. The error dynamics can be expanded as:
\begin{align} \label{eq:error_gfl_ineq}
\|\bm{e}_i(k+1)\| &= \|\mathscr{P}_{\mathcal{X}_i}(\bm{v}_i(k)) - \Delta \bm{x}^{ref}(k+1)\| \nonumber \\
&= \|\mathscr{P}_{\mathcal{X}_i}(\bm{v}_i(k)) - \mathscr{P}_{\mathcal{X}_i}(\Delta \bm{x}^{ref}(k)) + \bm{\delta}_{i}(k)\| \nonumber \\
&\le \|\bm{v}_i(k) - \Delta \bm{x}^{ref}(k)\| + \|\bm{\delta}_{i}(k)\|,
\end{align}
where we utilize Lemma \ref{lem:projection} and introduce the perturbation term $\bm{\delta}_{i}(k) = \mathscr{P}_{\mathcal{X}_i}(\Delta \bm{x}^{ref}(k)) - \Delta \bm{x}^{ref}(k+1)$. This term accounts for two effects: the drift of the reference due to dynamic loads (Assumption \ref{asm:bounded_rate}) and the potential mismatch if the ideal reference momentarily exceeds local constraints. Both effects are bounded by design, i.e., $\|\bm{\delta}_{i}(k)\| \le \xi_{local}$.

\begin{figure}[!t]
    \centering
    \includegraphics[width=0.75\columnwidth]{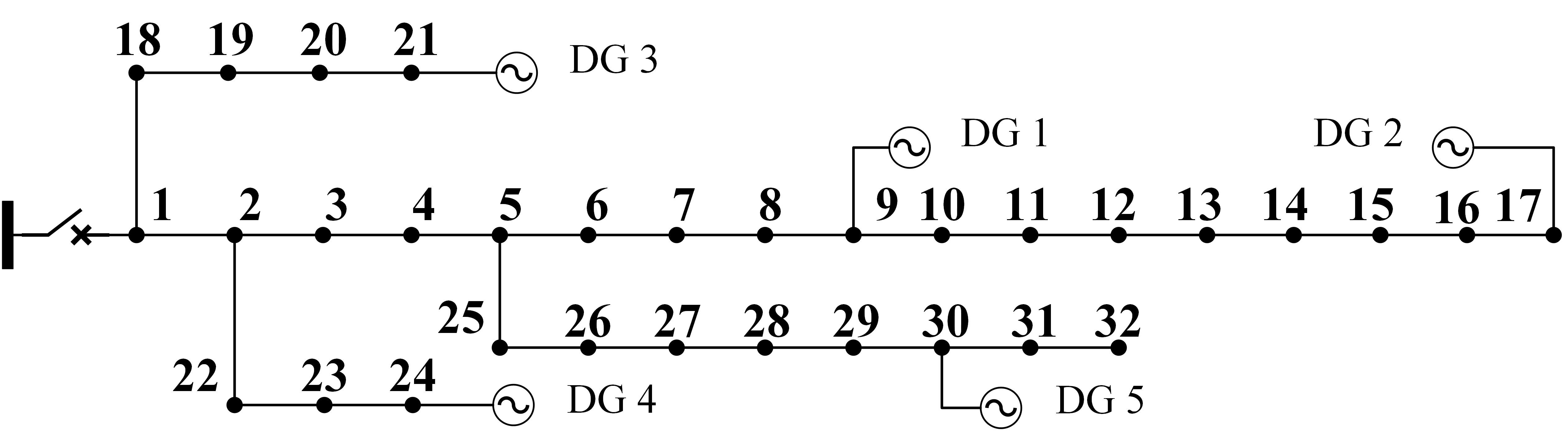}
    \caption{Topology of the modified IEEE 33-bus hybrid microgrid system.}
    \label{fig:ieee33}
\end{figure}

\begin{figure}[!t]
    \centering
    \includegraphics[width=0.5\columnwidth]{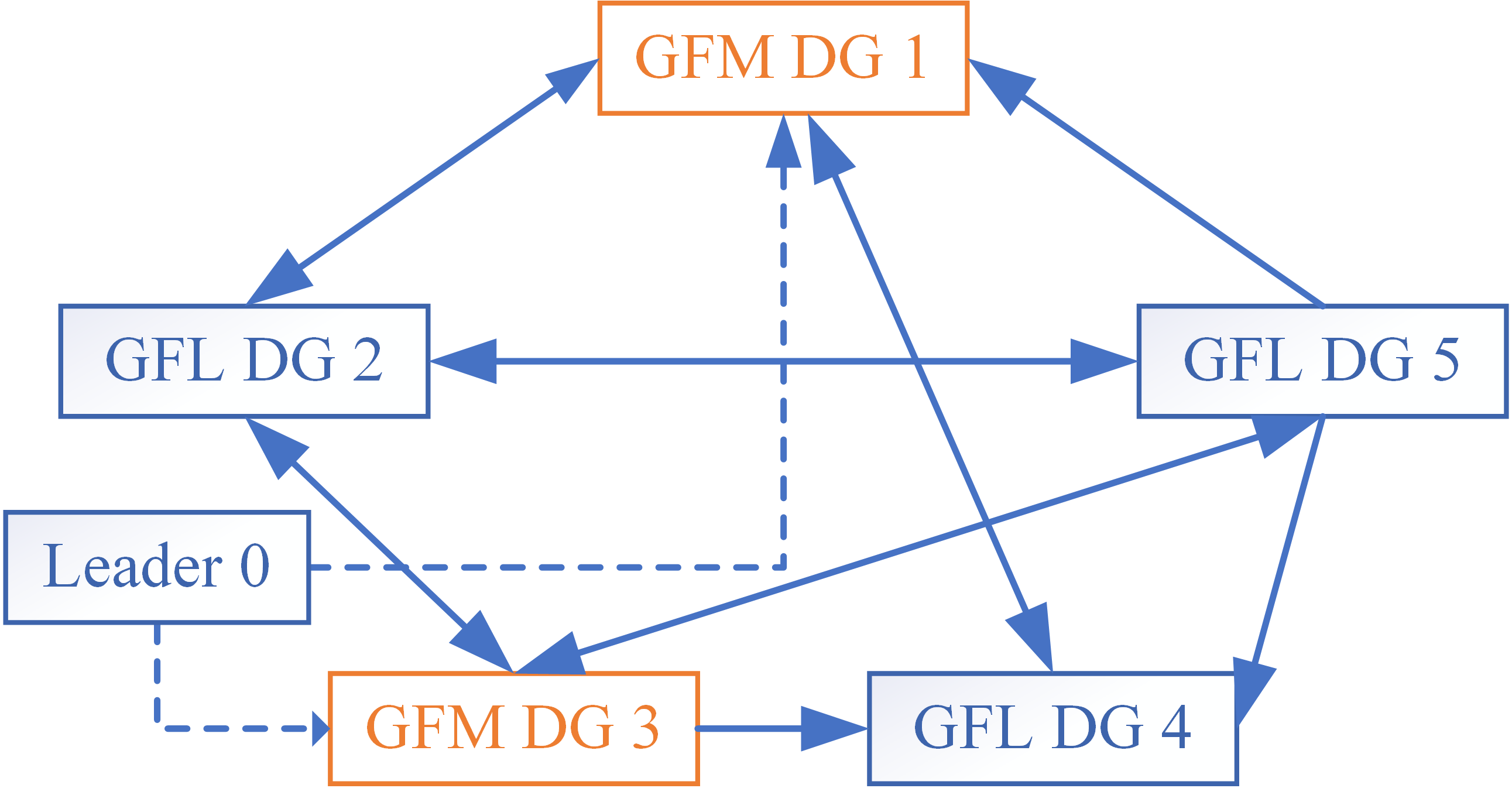}
    \caption{Communication topology for the hybrid GFL/GFM system.}
    \label{fig:comm_topo}
\end{figure}

\begin{table}[!t]
\centering
\caption{DG Parameters for the IEEE 33-bus Microgrid}
\label{tab:dg_params}
\begin{tabular}{cccccc}
\hline
DG& Type & $P_{DG}$ & Limits & $Q_{DG}$ & Limits \\
 ID& & (MW)& (MW)& (MVar)&(MVar)\\ \hline
1, 3 & GFM & 0.8 & $(-\infty, \infty)$ & 0.6 & $(-\infty, \infty)$ \\
2, 4 & GFL & 0.4 & $(0.3, 0.5)$ & 3.0 & $(0, 0.35)$ \\
5 & GFL & 0.4 & $(0, 0.8)$ & 3.0 & $(-0.6, 0.6)$ \\ \hline
\end{tabular}
\end{table}

\begin{figure}[!t]
    \centering
    \includegraphics[width=0.5\columnwidth]{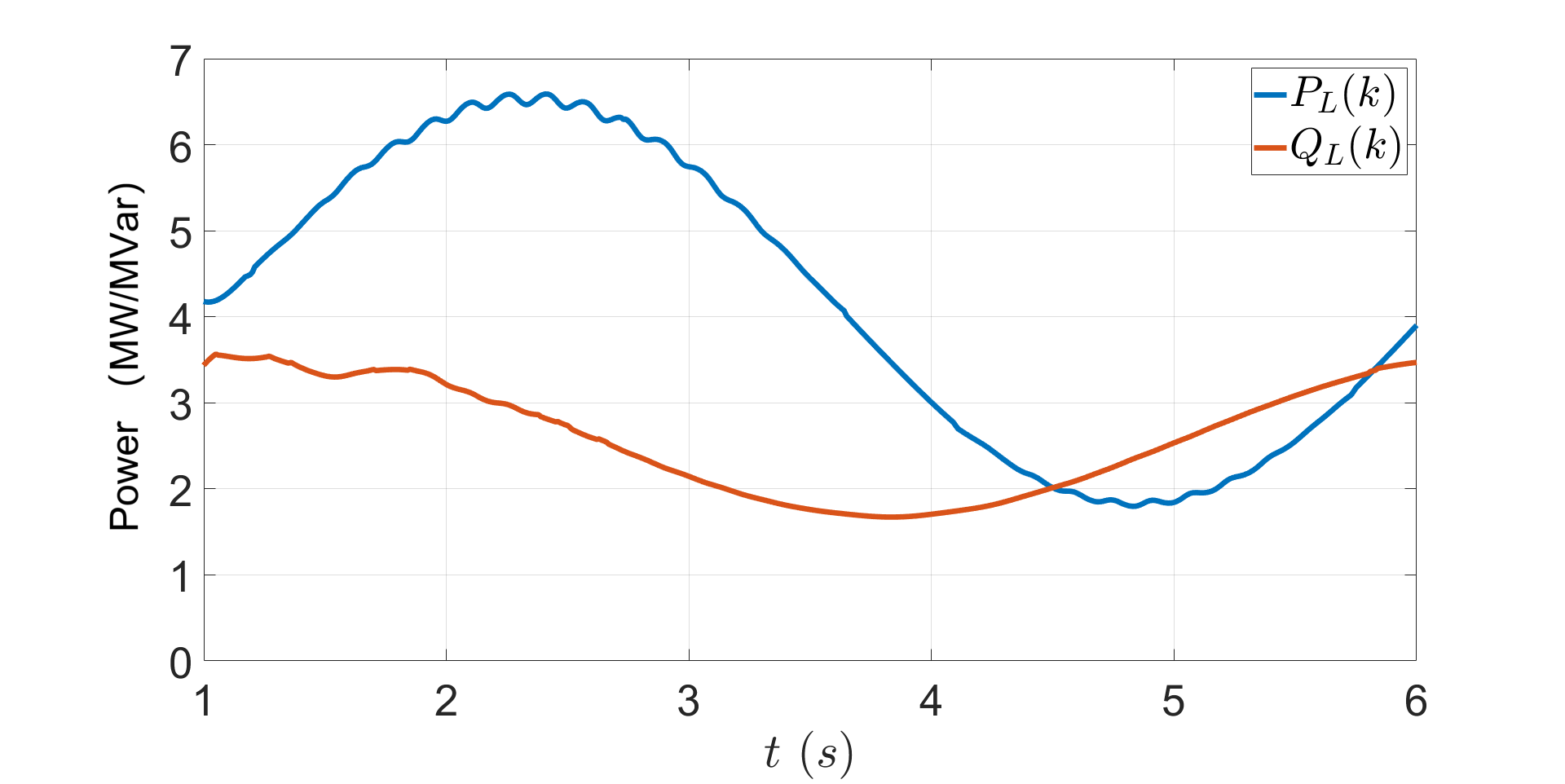}
    \caption{Total active and reactive load fluctuations in grid-connected mode.}
    \label{fig:grid_load}
\end{figure}

\begin{figure*}[t]
    \centering
    \subfloat[Active Power Output]{\includegraphics[width=0.44\linewidth]{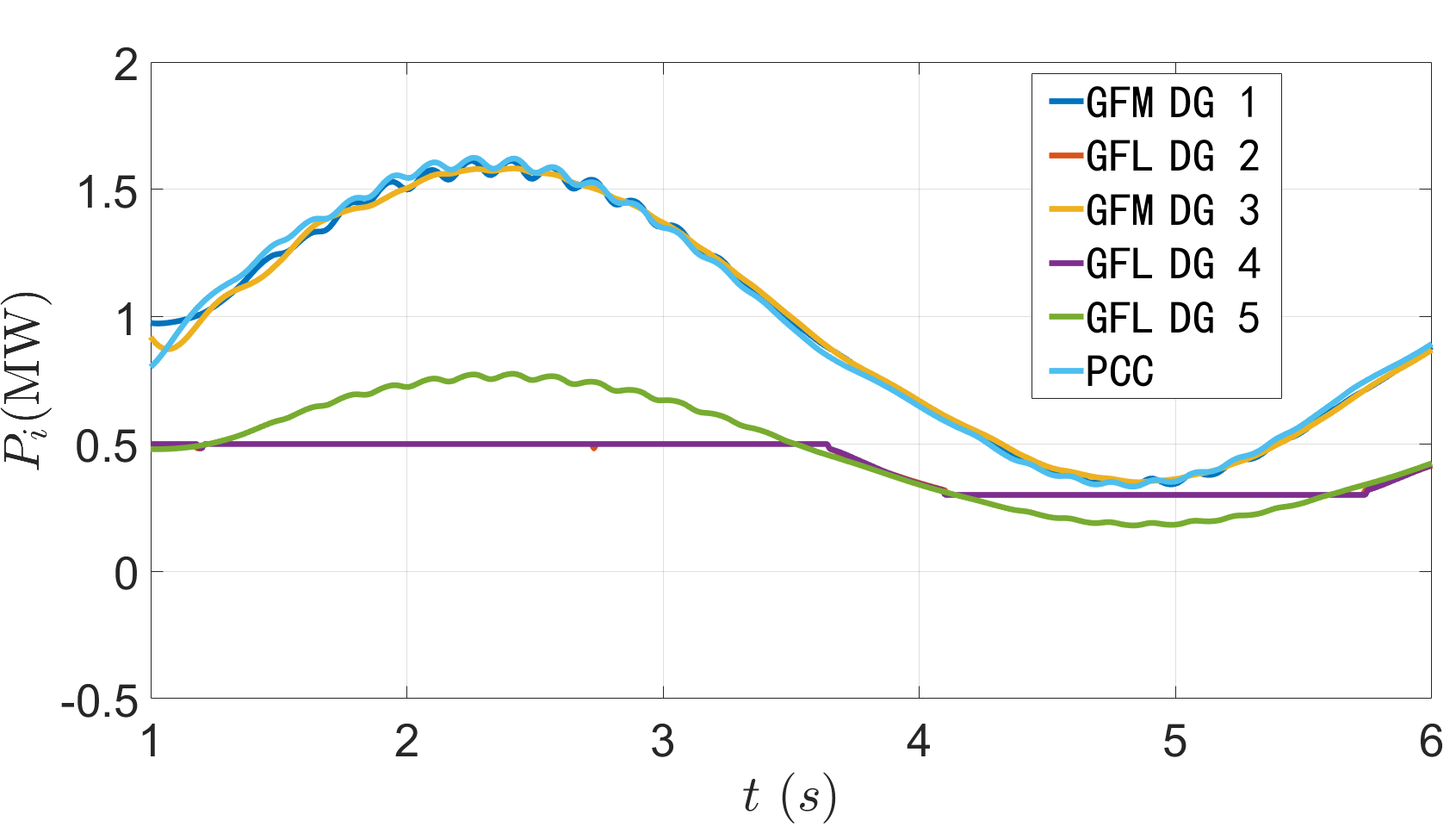}\label{fig:grid_withac_P}}
    \hfill
    \subfloat[Standardized Active Increment]{\includegraphics[width=0.44\linewidth]{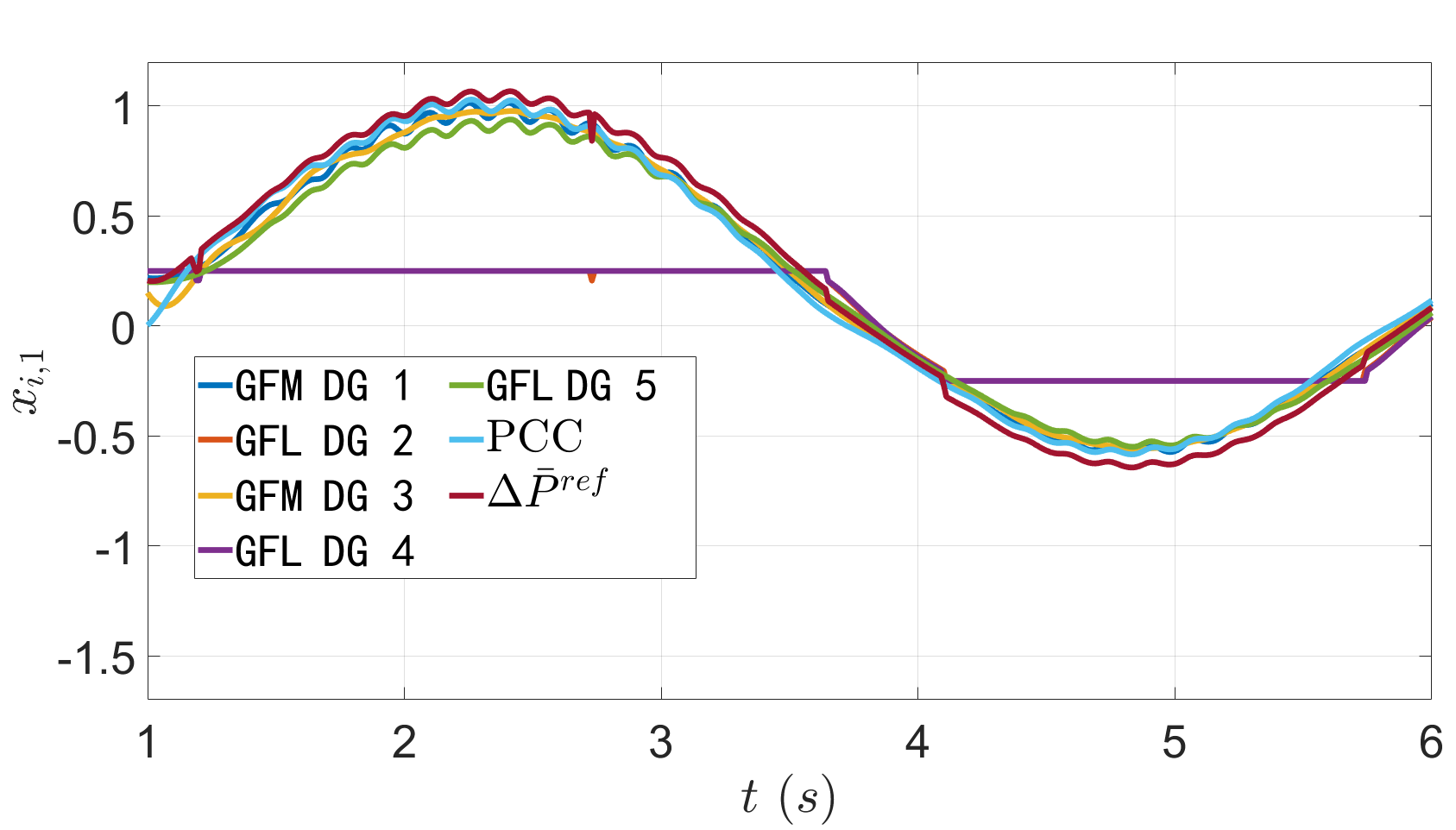}\label{fig:grid_withac_Pd}}
    \\
    \subfloat[Reactive Power Output]{\includegraphics[width=0.44\linewidth]{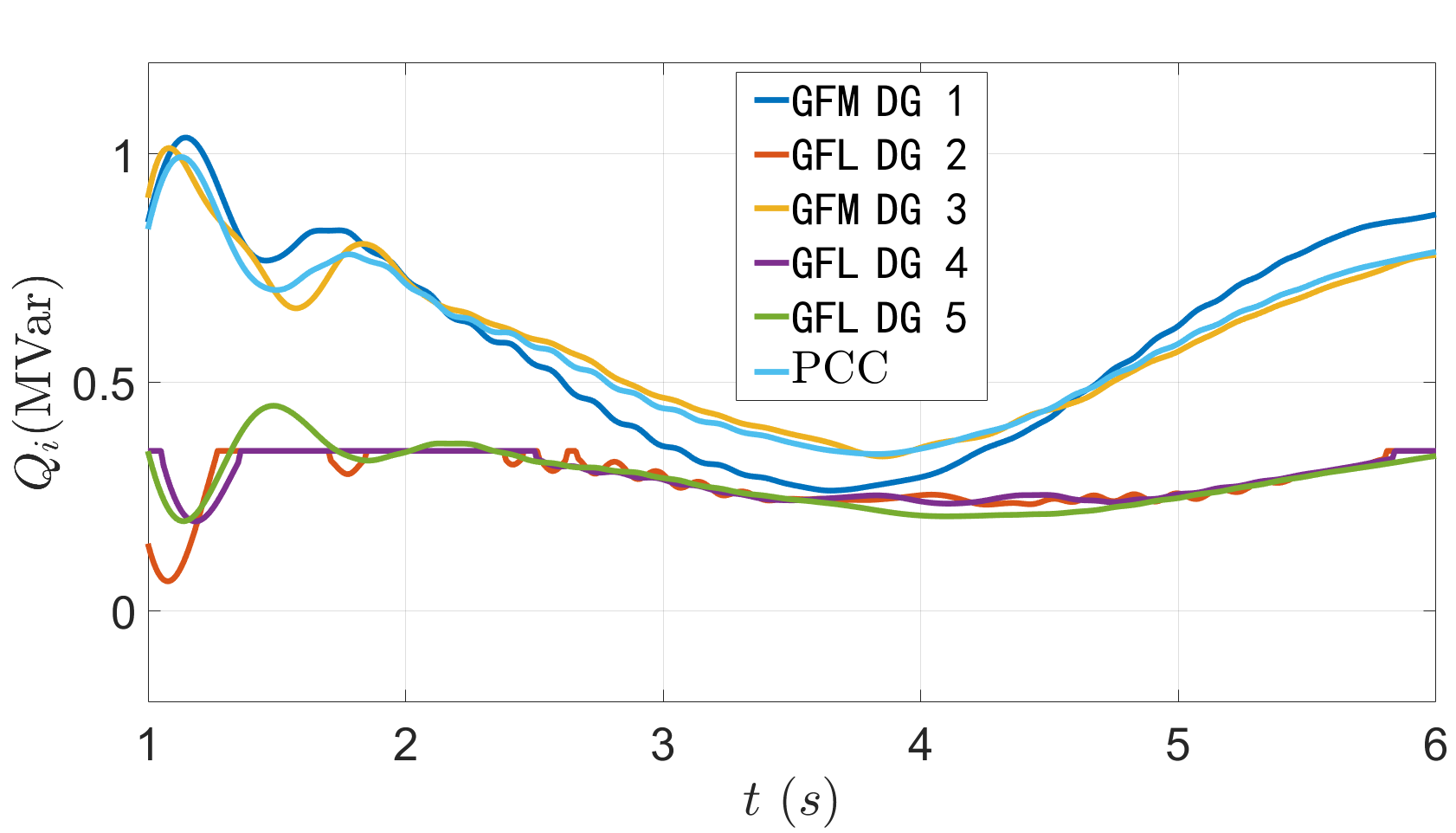}\label{fig:grid_withac_Q}}
    \hfill
    \subfloat[Standardized Reactive Increment]{\includegraphics[width=0.44\linewidth]{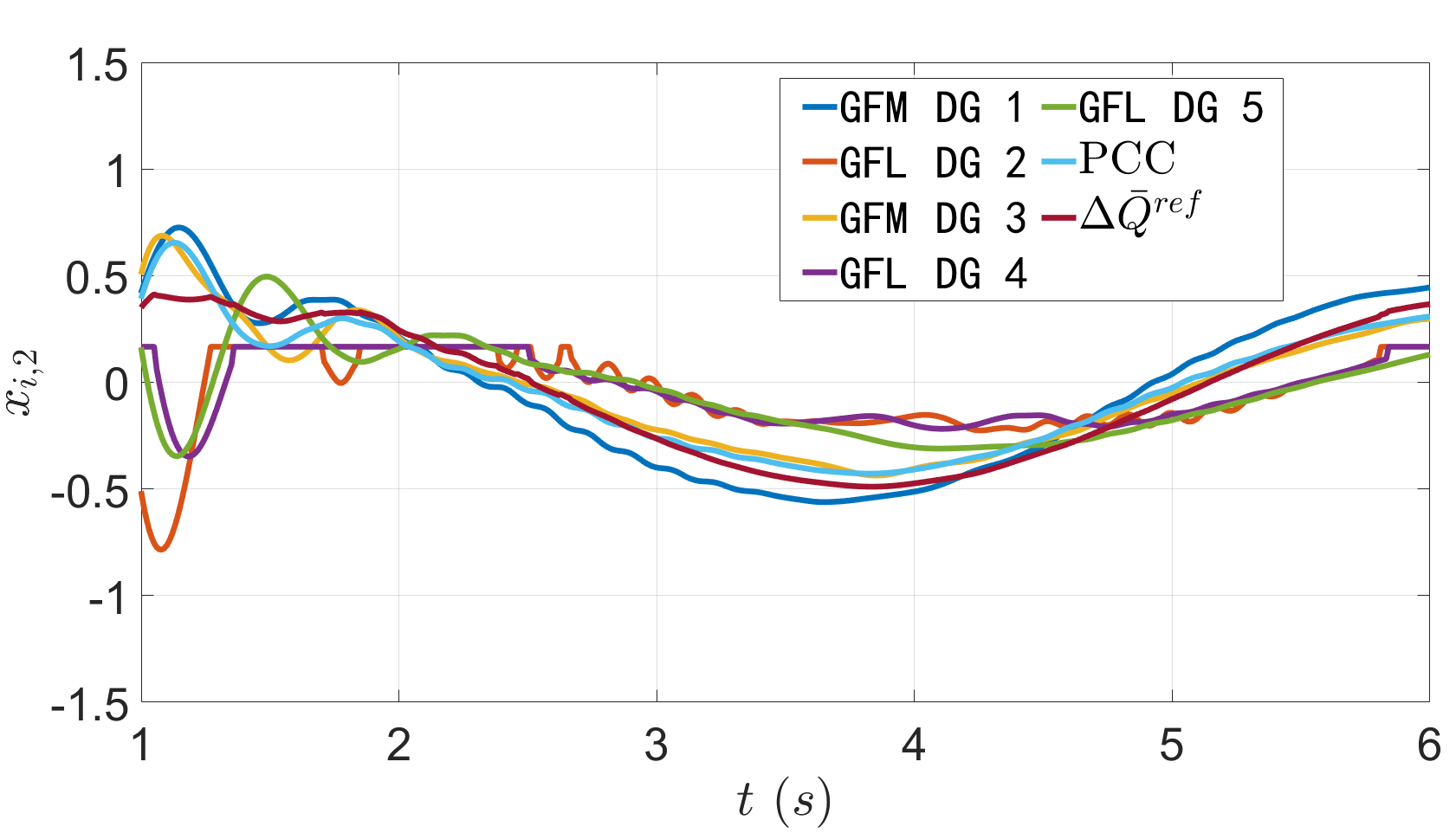}\label{fig:grid_withac_Qd}}
    \caption{Dynamic response in grid-connected mode using the \textbf{Proposed} strategy (With Activation Function).}
    \label{fig:grid_withac}
\end{figure*}

\begin{figure*}[t]
    \centering
    \subfloat[Active Power Output]{\includegraphics[width=0.44\linewidth]{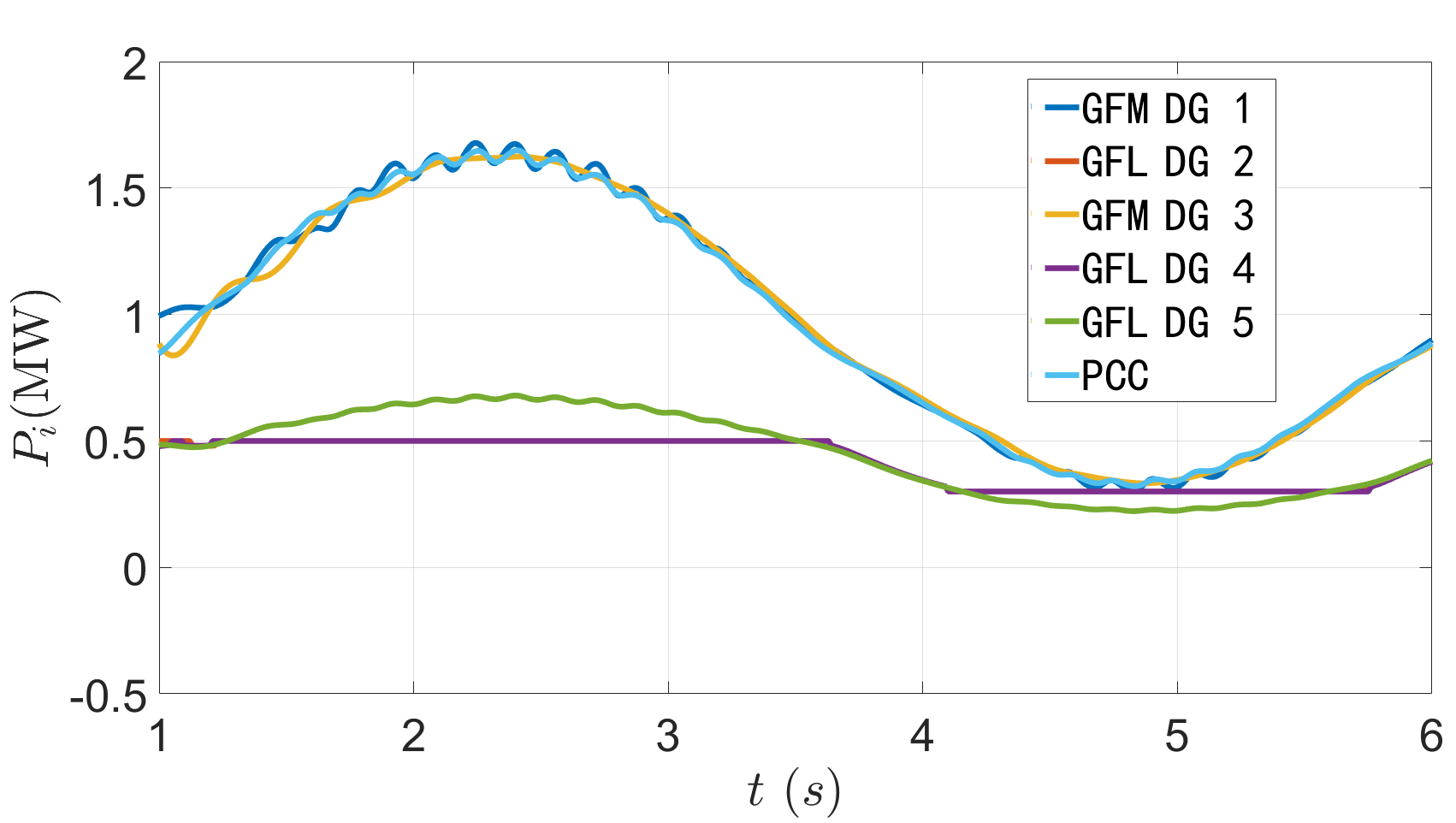}\label{fig:grid_withoutac_P}}
    \hfill
    \subfloat[Standardized Active Increment]{\includegraphics[width=0.44\linewidth]{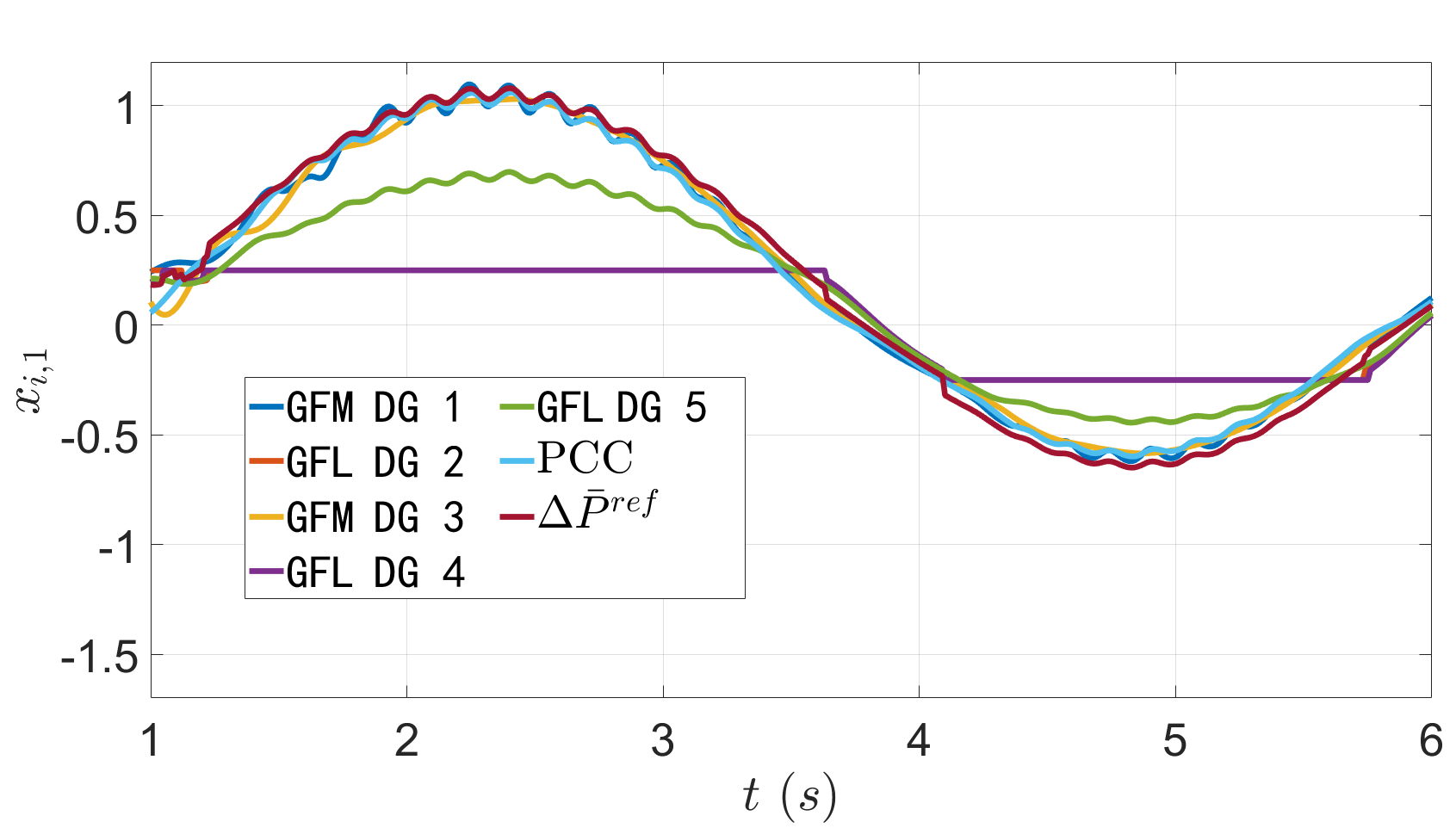}\label{fig:grid_withoutac_Pd}}
    \\
    \subfloat[Reactive Power Output]{\includegraphics[width=0.44\linewidth]{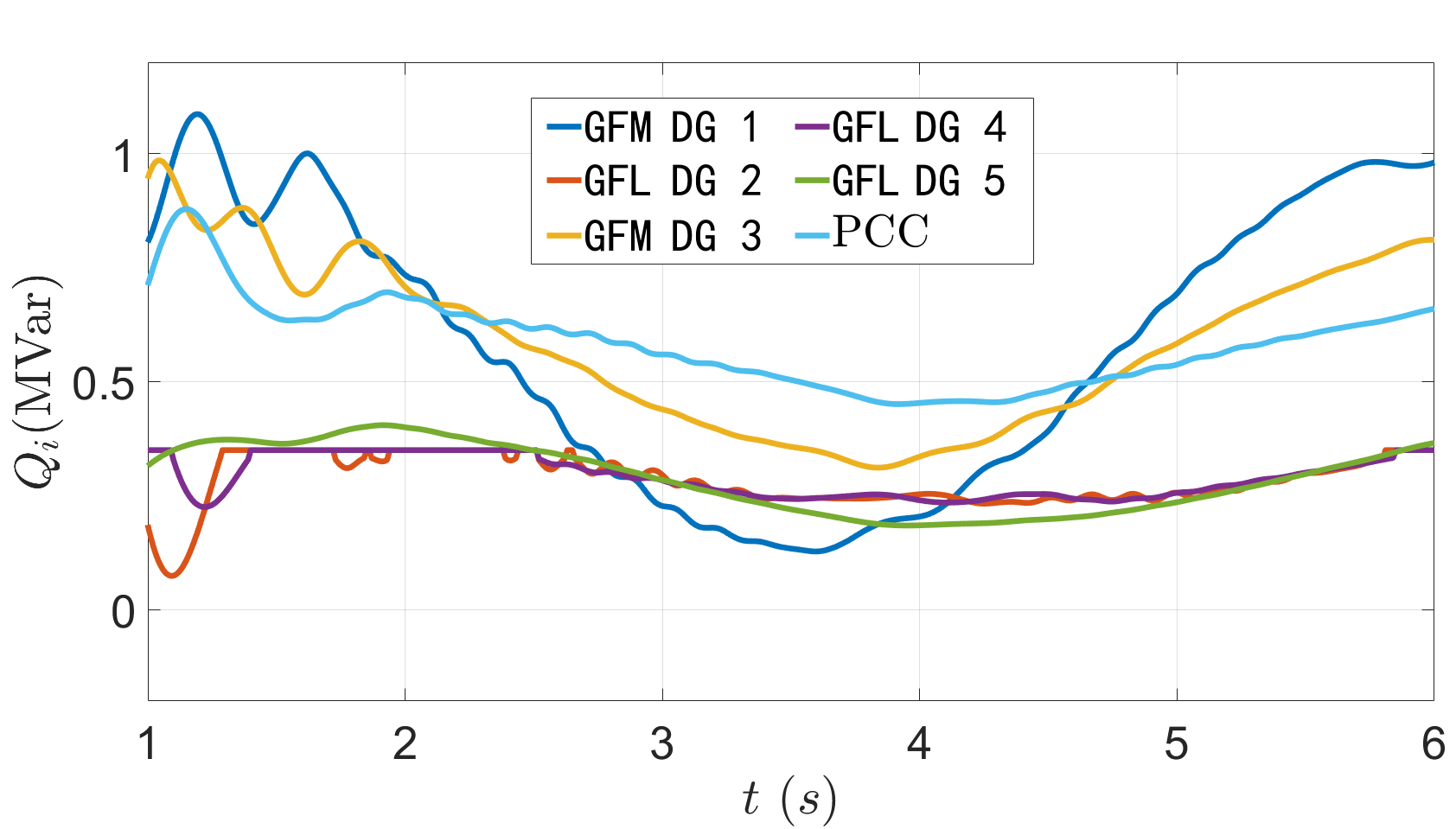}\label{fig:grid_withoutac_Q}}
    \hfill
    \subfloat[Standardized Reactive Increment]{\includegraphics[width=0.44\linewidth]{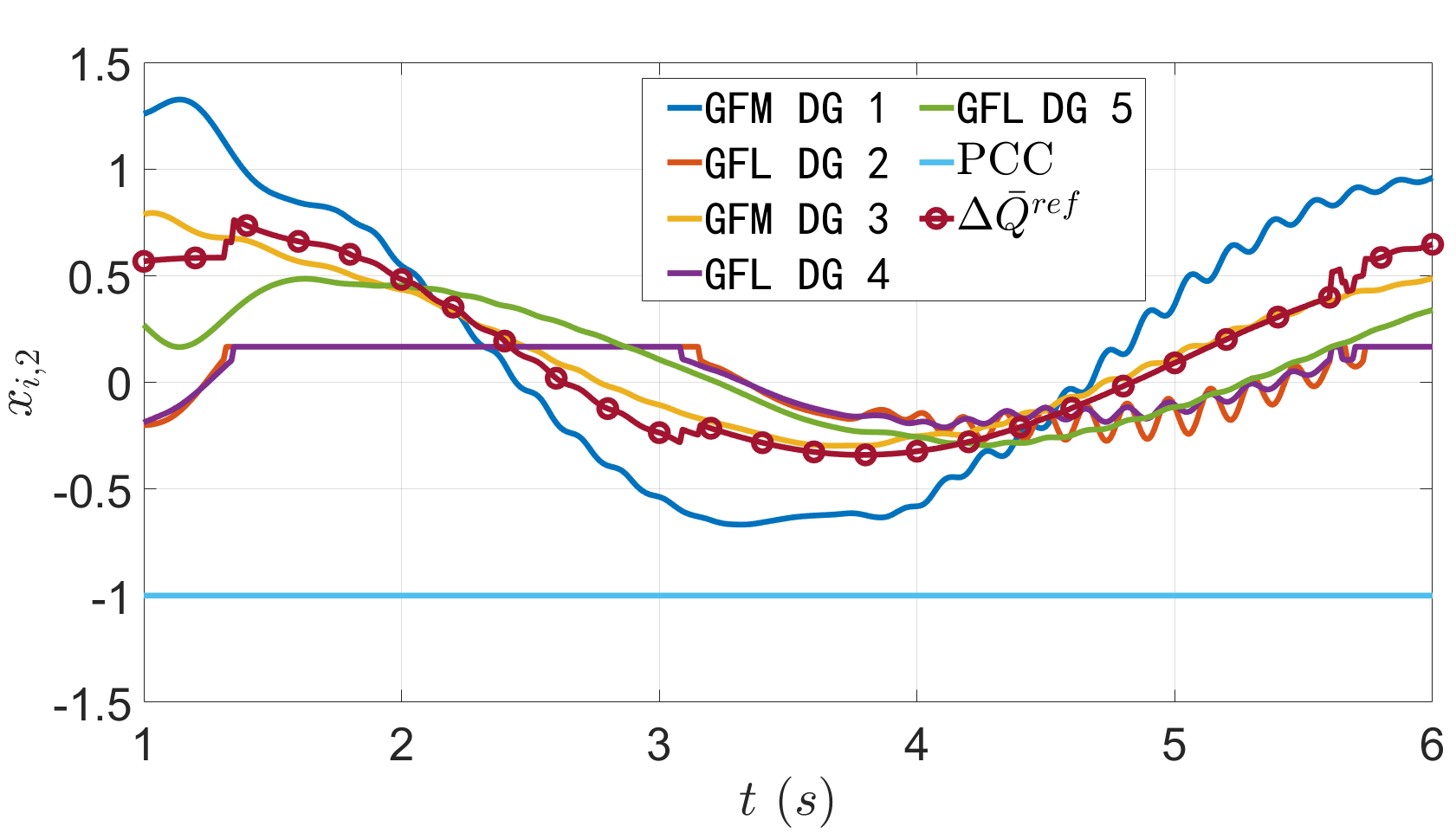}\label{fig:grid_withoutac_Qd}}
    \caption{Dynamic response in grid-connected mode using the \textbf{Traditional} strategy (Without Activation Function).}
    \label{fig:grid_withoutac}
\end{figure*}

\begin{figure}[t]
    \centering
    \includegraphics[width=0.5\columnwidth]{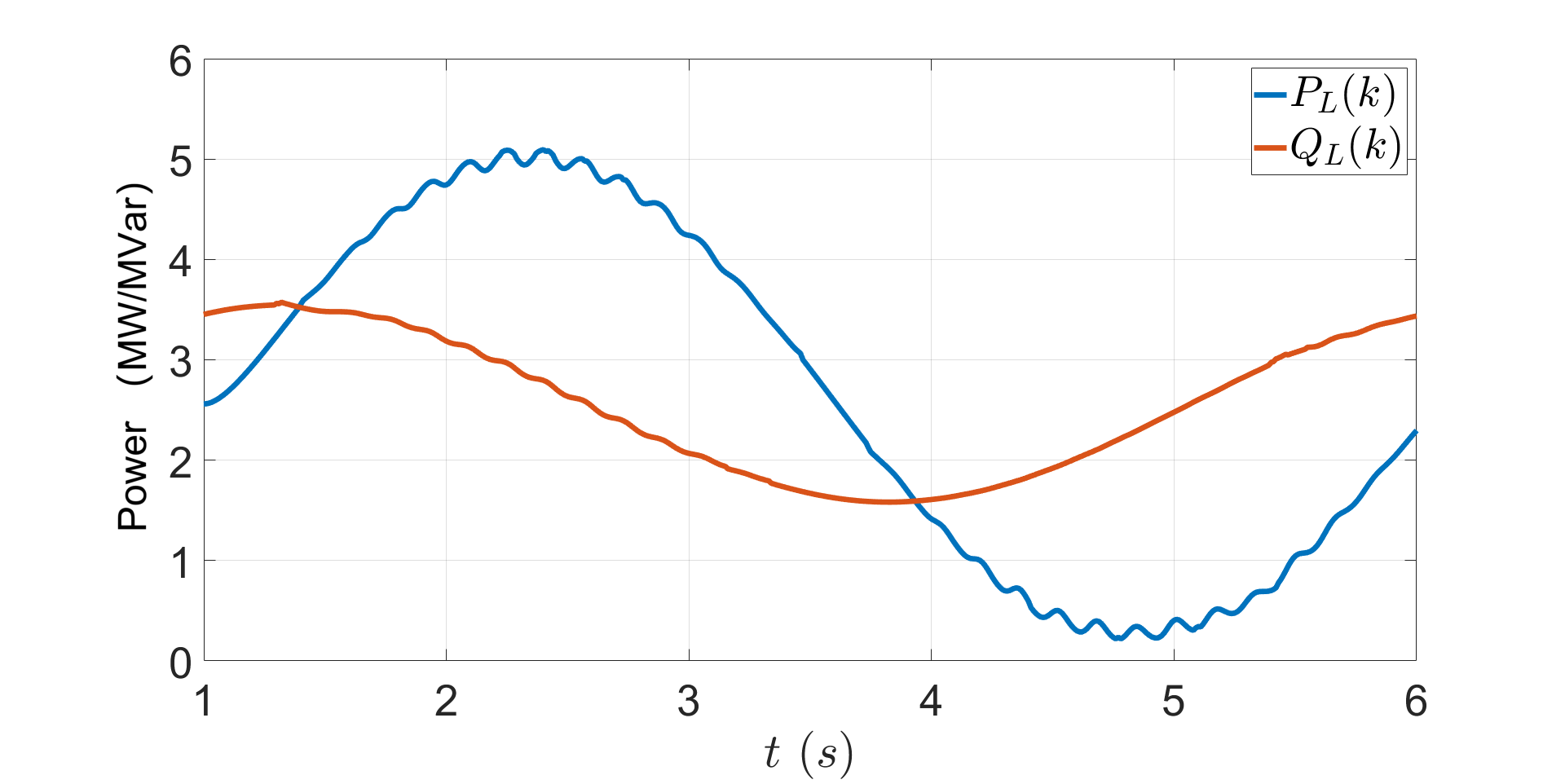}
    \caption{Total active and reactive load fluctuations in islanded mode.}
    \label{fig:island_load}
\end{figure}

\begin{figure*}[t]
    \centering
    \subfloat[Active Power Output]{\includegraphics[width=0.44\linewidth]{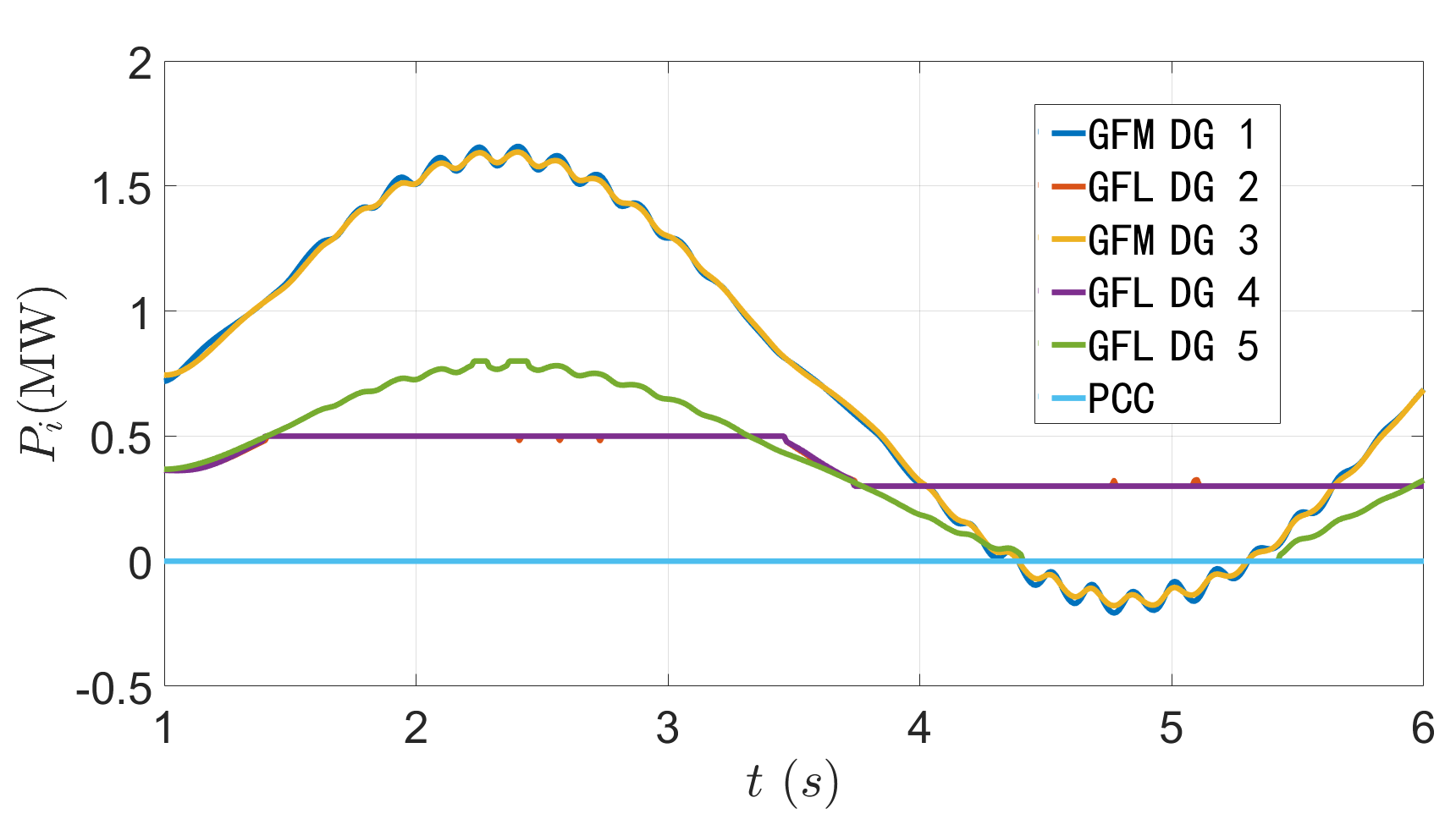}\label{fig:island_withac_P}}
    \hfill
    \subfloat[Standardized Active Increment]{\includegraphics[width=0.44\linewidth]{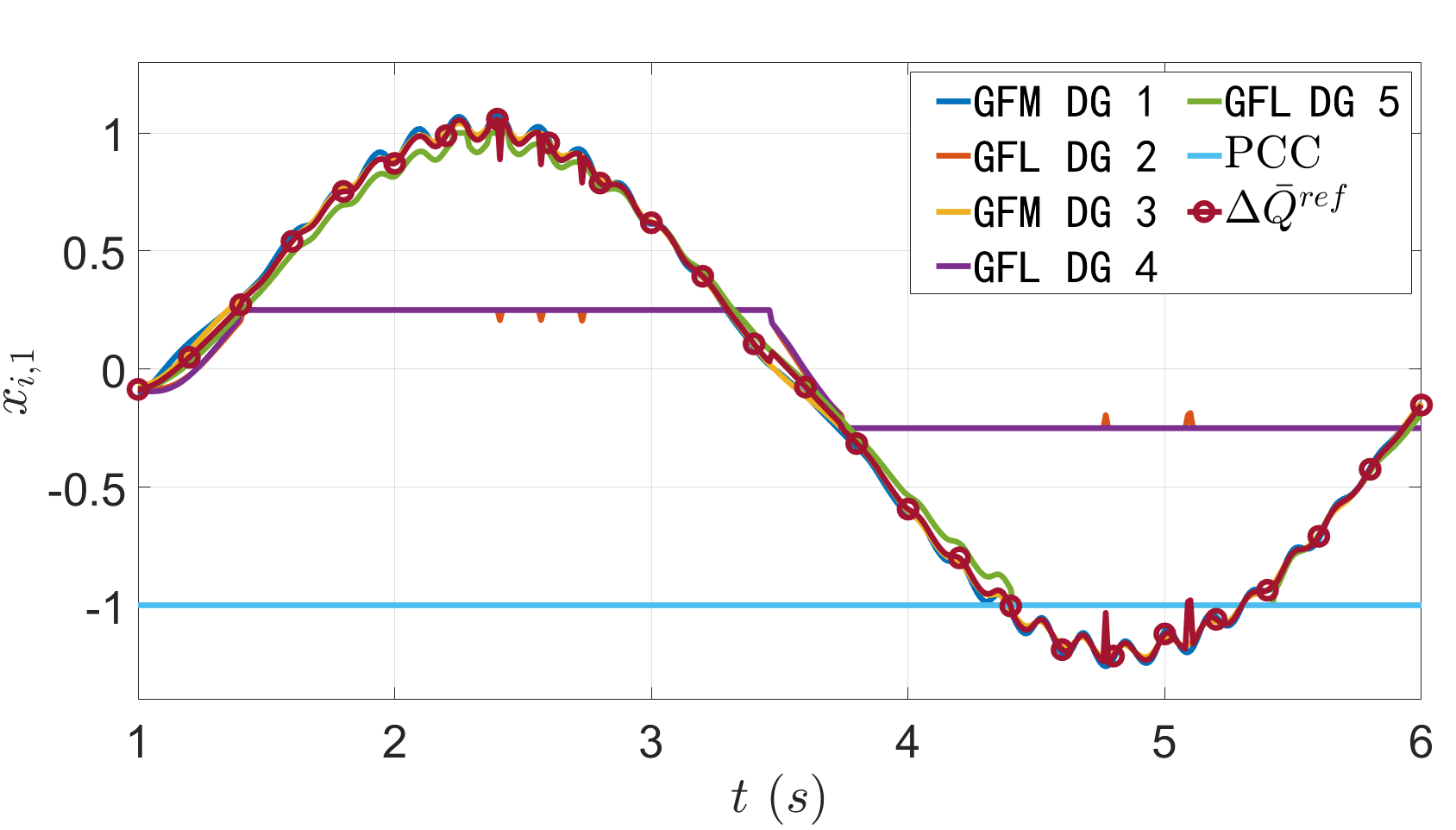}\label{fig:island_withac_Pd}}
    \\
    \subfloat[Reactive Power Output]{\includegraphics[width=0.44\linewidth]{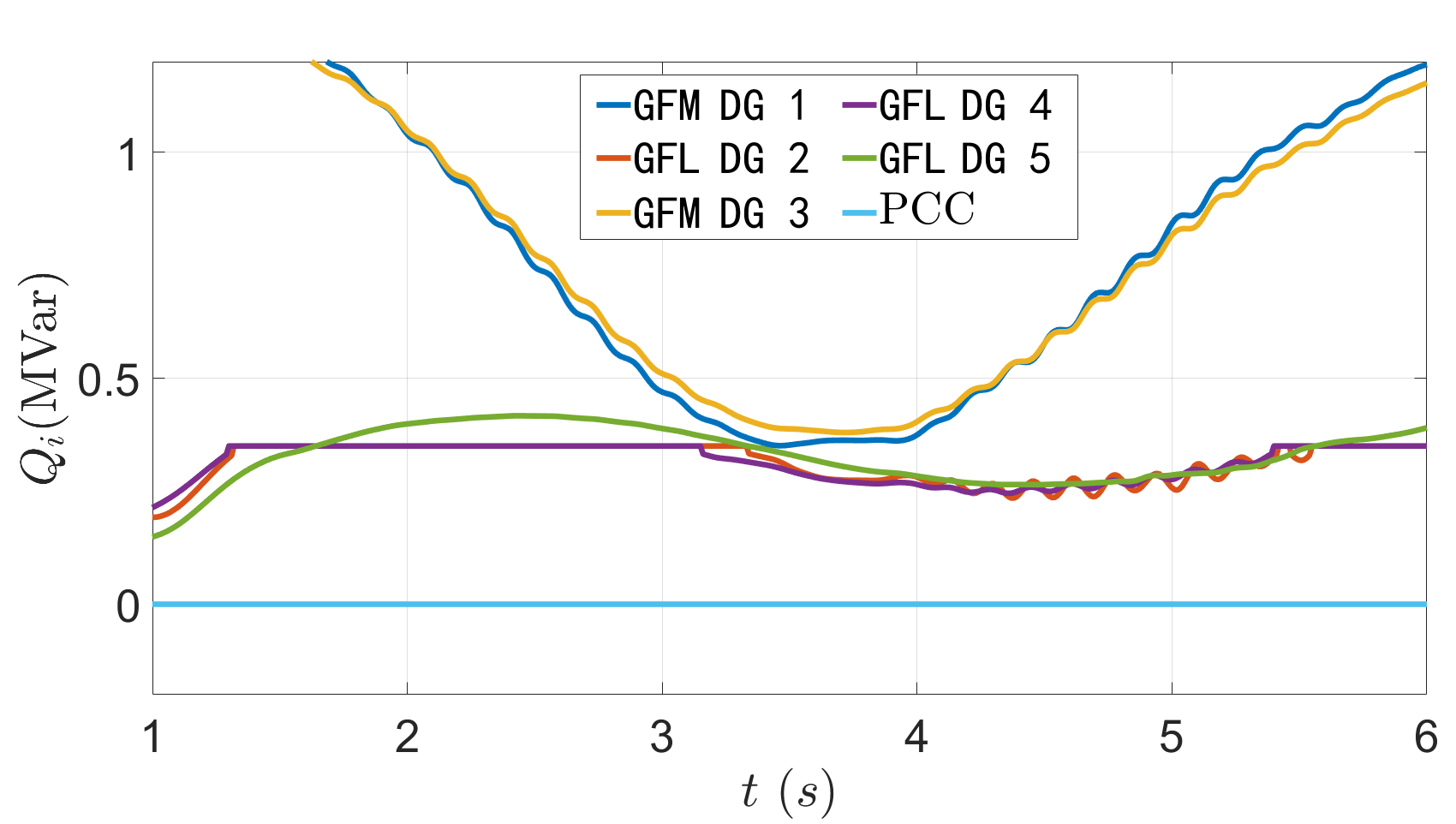}\label{fig:island_withac_Q}}
    \hfill
    \subfloat[Standardized Reactive Increment]{\includegraphics[width=0.44\linewidth]{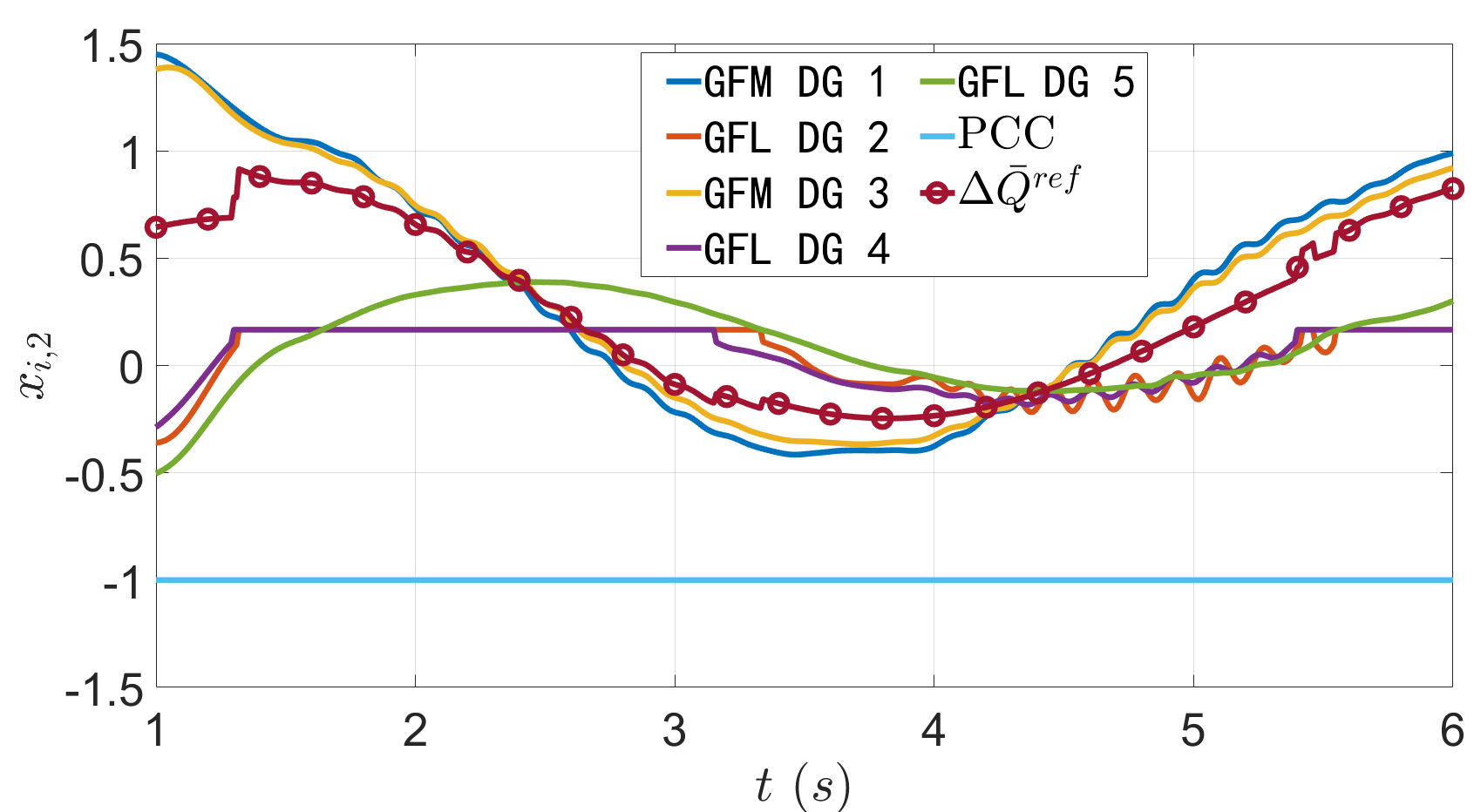}\label{fig:island_withac_Qd}}
    \caption{Dynamic response in islanded mode using the \textbf{Proposed} strategy (With Activation Function).}
    \label{fig:island_withac}
\end{figure*}
\begin{figure*}[t]
    \centering
    \subfloat[Active Power Output]{\includegraphics[width=0.44\linewidth]{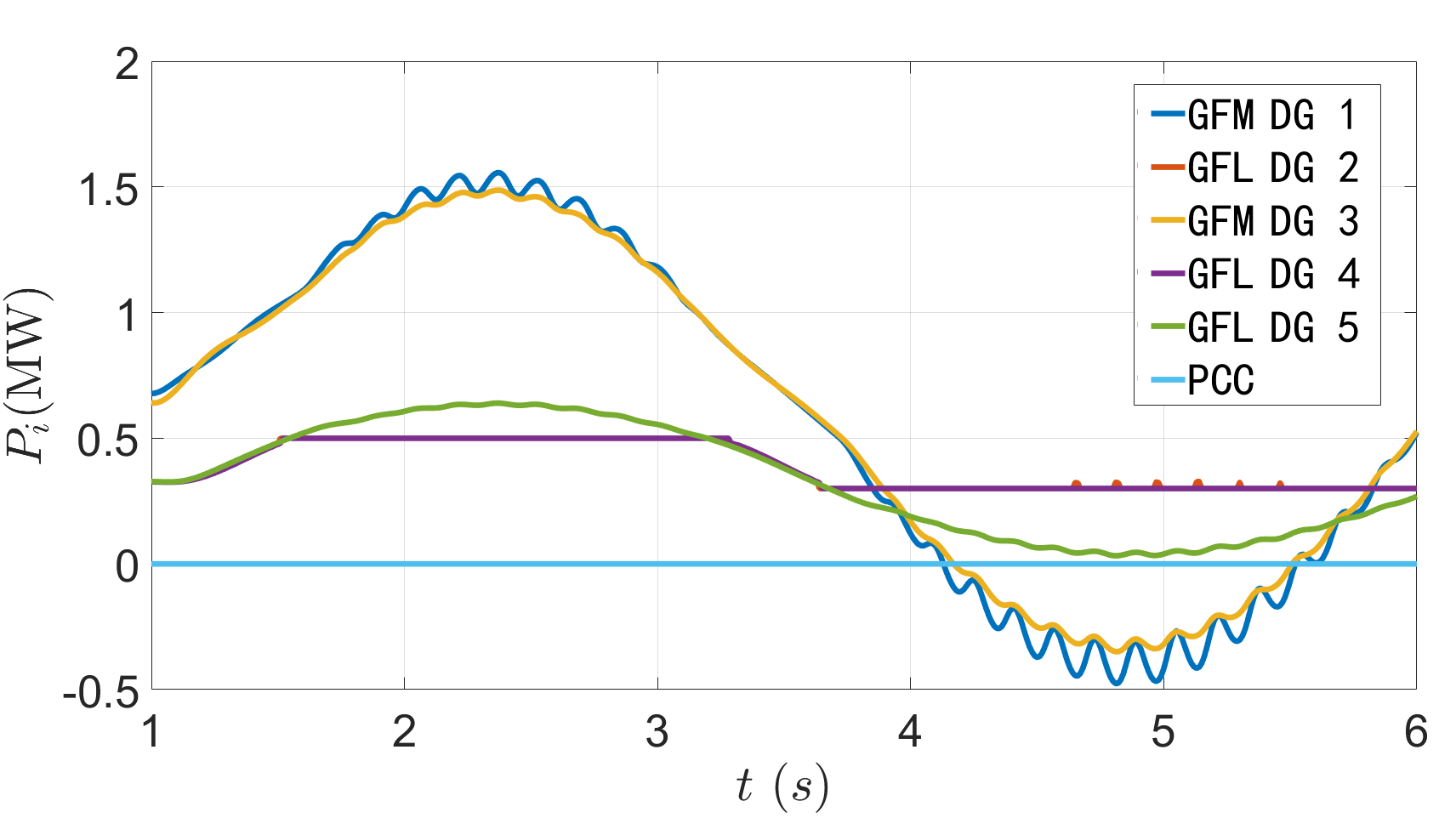}\label{fig:island_withoutac_P}}
    \hfill
    \subfloat[Standardized Active Increment]{\includegraphics[width=0.44\linewidth]{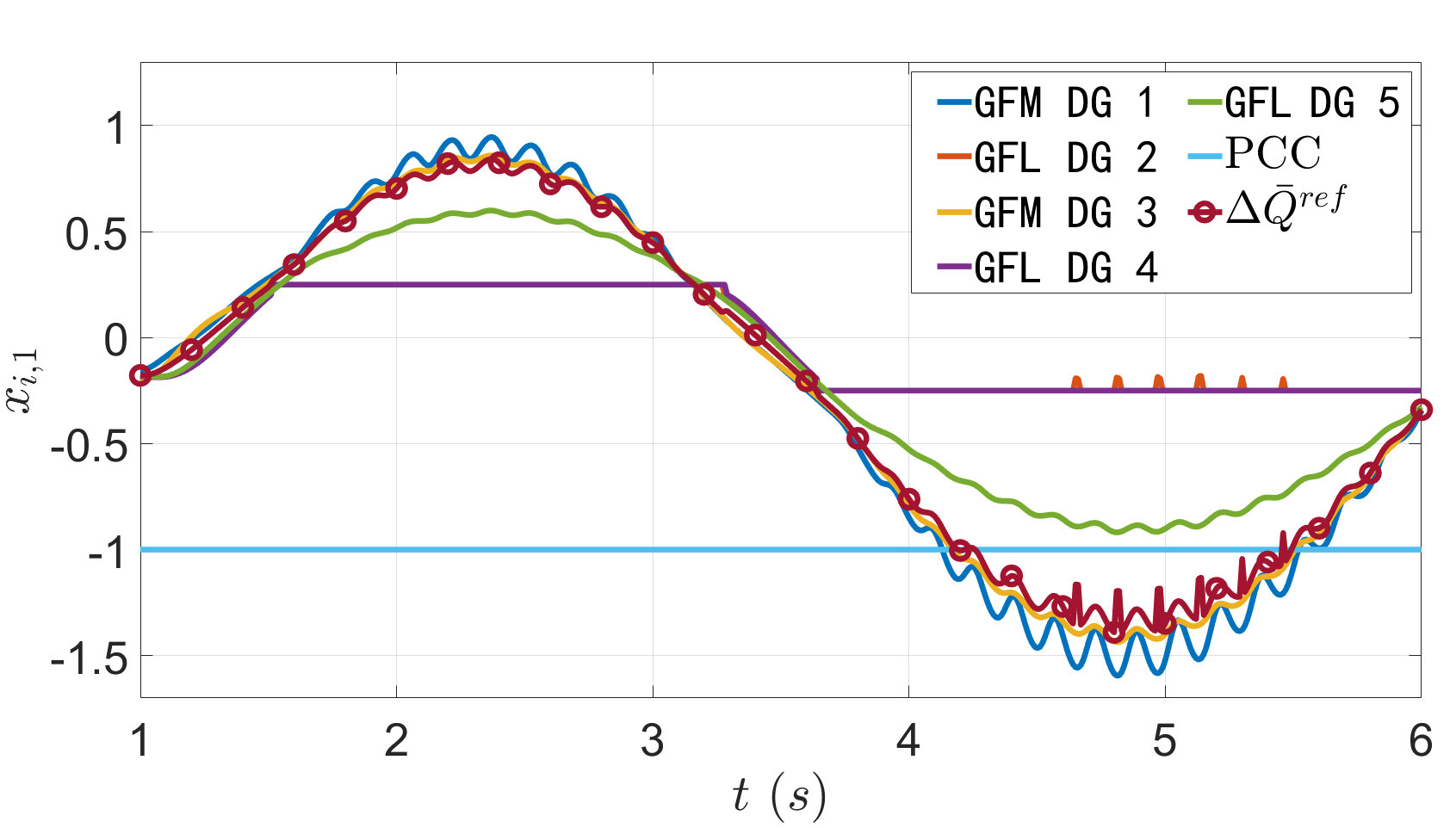}\label{fig:island_withoutac_Pd}}
    \\
    \subfloat[Reactive Power Output]{\includegraphics[width=0.44\linewidth]{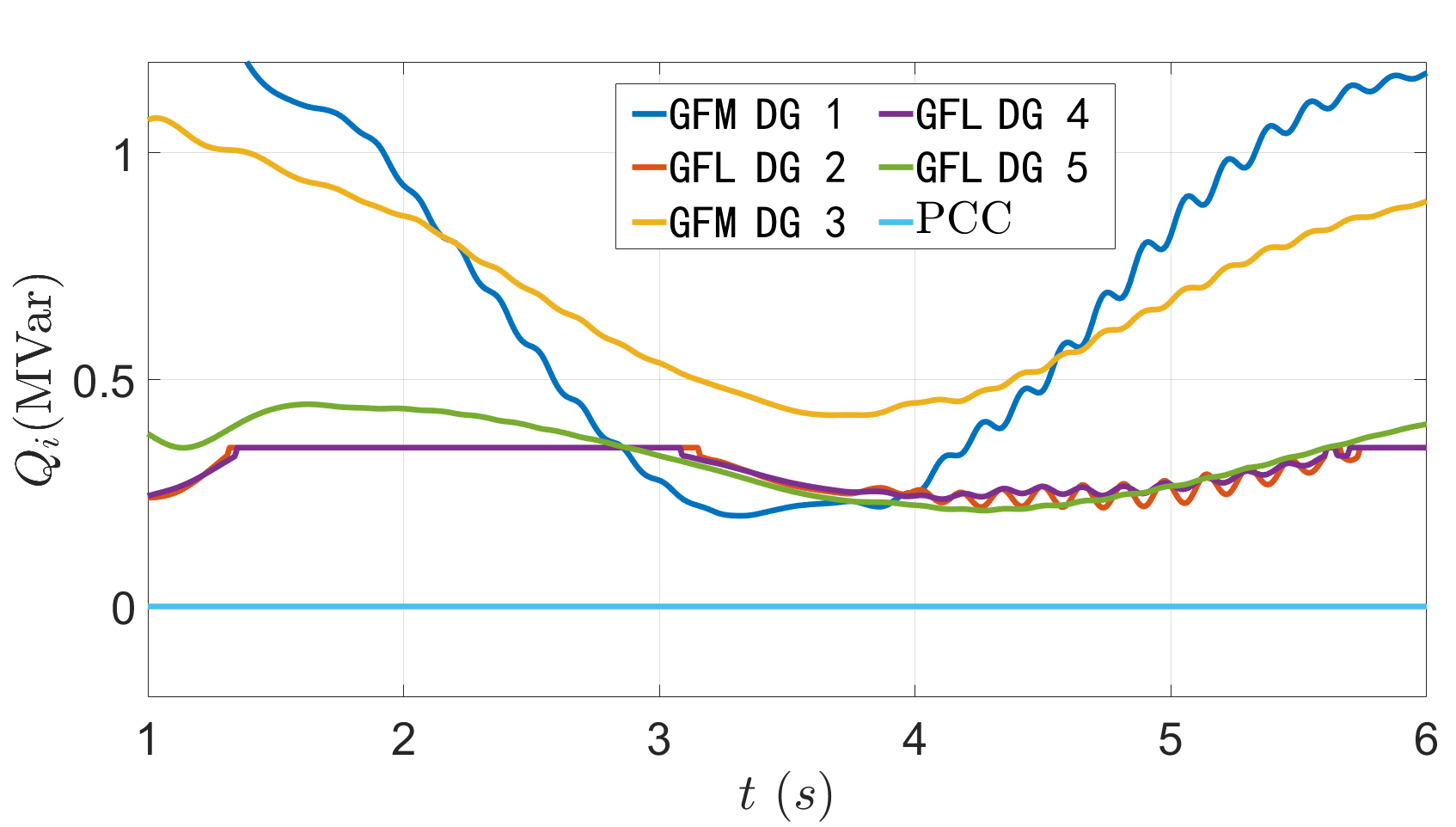}\label{fig:island_withoutac_Q}}
    \hfill
    \subfloat[Standardized Reactive Increment]{\includegraphics[width=0.44\linewidth]{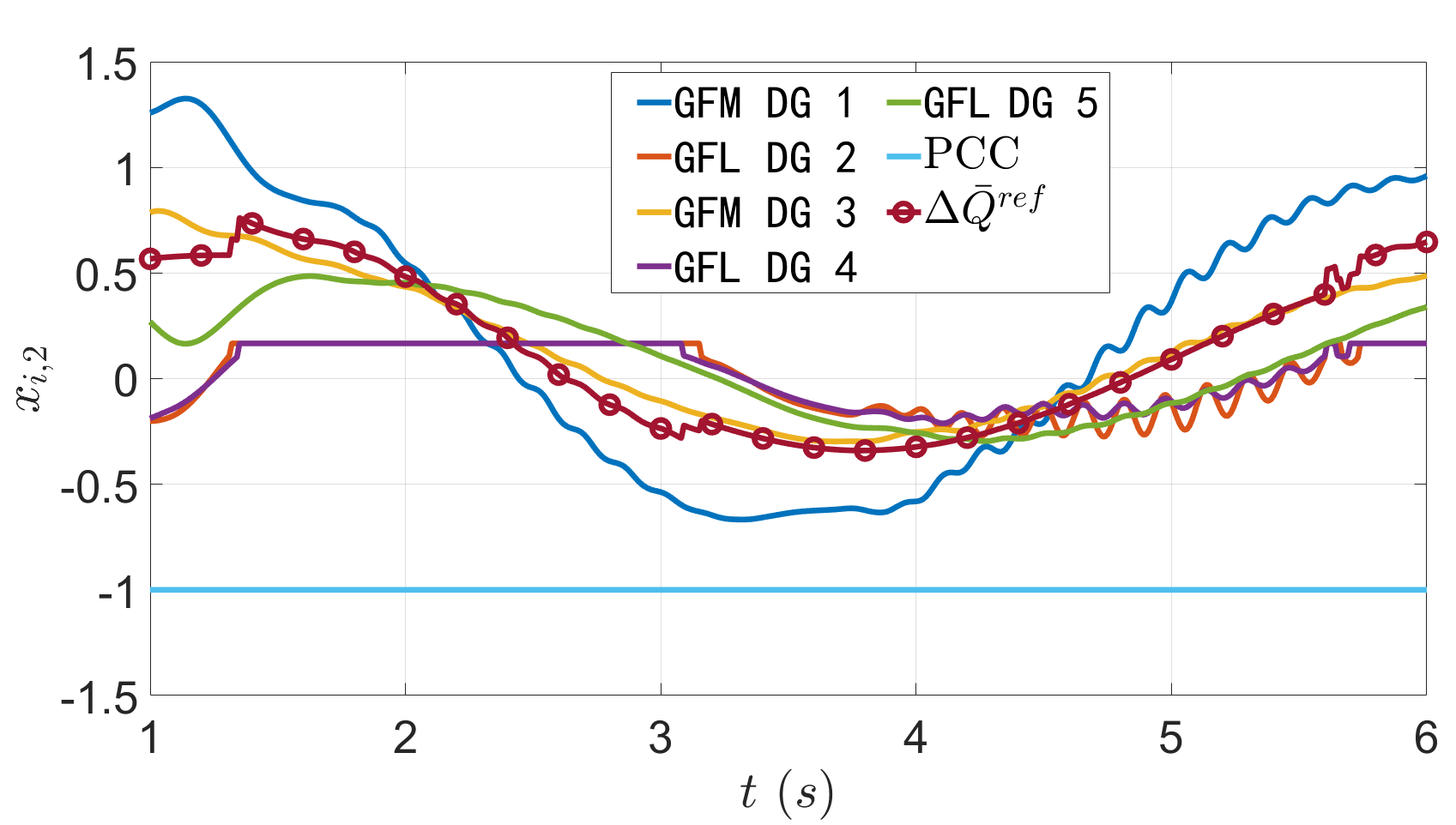}\label{fig:island_withoutac_Qd}}
    \caption{Dynamic response in islanded mode using the \textbf{Traditional} strategy (Without Activation Function).}
    \label{fig:island_withoutac}
\end{figure*}

\subsubsection{Global Matrix Formulation}
The global error dynamics can be formulated in a compact matrix form. The system matrix $\bm{W}_{\sigma(k)}$ is state-dependent. Specifically:
\begin{itemize}
    \item For unsaturated nodes (GFM and active GFL), the rows of $\bm{W}_{\sigma(k)}$ correspond to a standard consensus update with weights $c \cdot a_{ij}$.
    \item For saturated GFL nodes, the effective update becomes $\bm{x}_i(k+1) \approx \bm{x}_i(k)$ (maintained at boundary), implying the corresponding row in $\bm{W}_{\sigma(k)}$ approaches the identity vector, but decoupled from neighbors.
\end{itemize}
Consequently, the error dynamics are:
\begin{equation} \label{eq:global_error}
\bm{E}(k+1) = \bm{W}_{\sigma(k)} \bm{E}(k) + \bm{\Delta}_{total}(k),
\end{equation}
where $\bm{\Delta}_{total}(k)$ aggregates the load drift, attack residuals, and projection errors.

\subsubsection{Convergence Analysis of Switching System}
The stability hinges on the spectral properties of the set of matrices $\{\bm{W}_{\sigma}\}$. Under Assumption \ref{asm:topology}, for any possible activation configuration $\sigma$, the subgraph formed by the virtual leader (SCADA) and the unsaturated GFM nodes retains a spanning tree rooted at the leader. The saturated GFL nodes, being logically disconnected by $\bm{A}^{ac}$, act as bounded exogenous inputs to the rest of the system rather than destabilizing feedback loops.

It is a standard result in switching system theory that if all matrices in the switching set share a Common Quadratic Lyapunov Function (CQLF) or if the union of graphs is sufficiently connected, the system is stable. Here, the presence of the GFM backbone ensures that the spectral radius of the essential part of $\bm{W}_{\sigma(k)}$ (excluding the saturated nodes) is strictly less than unity. Let $\rho = \sup_{\sigma} \rho(\bm{W}_{\sigma}^{essential}) < 1$.
Taking the norm of the error system:
\begin{align}
\|\bm{E}(k+1)\| &\le \|\bm{W}_{\sigma(k)}\| \|\bm{E}(k)\| + \|\bm{\Delta}_{total}(k)\| \nonumber \\
&\le \rho \|\bm{E}(k)\| + \Xi,
\end{align}
where $\Xi = \sup_k \|\bm{\Delta}_{total}(k)\|$.

Since $0 \le \rho < 1$, the error dynamics represent a contraction mapping subject to a bounded disturbance. By recursively applying this inequality, we obtain:
\begin{equation}
\|\bm{E}(k)\| \le \rho^k \|\bm{E}(0)\| + \Xi \sum_{j=0}^{k-1} \rho^j.
\end{equation}
As $k \to \infty$, the initial error term decays to zero, and the accumulated disturbance term converges to a finite limit:
\begin{equation}
\limsup_{k \to \infty} \|\bm{E}(k)\| \le \frac{\Xi}{1 - \rho} \triangleq \mu.
\end{equation}
This confirms that the synchronization error is Uniformly Ultimately Bounded. The bound $\mu$ is determined by the magnitude of load fluctuations (Assumption \ref{asm:bounded_rate}), the efficacy of attack mitigation (Lemma \ref{lem:attack_bound}), and the algebraic connectivity of the active network ($\rho$). 
\hfill $\blacksquare$

\vspace{-8pt}

\section{Case Studies}
\label{sec:sim}

To validate the effectiveness, resilience, and constraint-handling capability of the proposed hierarchical control strategy, comprehensive simulations are conducted on a modified IEEE 33-bus microgrid system. The case studies are designed to rigorously test the system under two distinct operational modes—grid-connected and islanded—while subjecting it to aggressive load fluctuations and mixed cyber-attacks. The primary objectives are to demonstrate: (1) the precise coordination between tertiary economic dispatch and secondary dynamic regulation via the standardized increment interface; (2) the necessity of the active constraint handling mechanism to prevent integrator wind-up during GFL saturation; and (3) the robustness of the system against unbounded FDI attacks and packet losses.

\vspace{-8pt}
\subsection{Experimental Setup and Parameters}

The test system is based on the IEEE 33-bus network topology (Fig. \ref{fig:ieee33}), modified to include a heterogeneous mix of GFL and GFM units. Specifically, DGs at buses 1 and 3 are configured as GFM units to provide voltage support, while DGs at buses 2, 4, and 5 operate as GFL units. The communication topology (Fig. \ref{fig:comm_topo}) satisfies the connectivity requirements of Assumption \ref{asm:topology}, ensuring a path exists from the GFM backbone to all GFL units.

The specific parameters for the DGs, including their tertiary power setpoints ($P_{DG}, Q_{DG}$) and physical output limits ($[\underline{P}, \overline{P}], [\underline{Q}, \overline{Q}]$), are detailed in Table \ref{tab:dg_params}. To rigorously test the cyber-resilience, the communication network is subjected to mixed faults: time-varying delays ($\tau \in [0, 20]$ms), packet loss probability ($\alpha = 0.1$), and unbounded FDI attacks on randomly selected links, consistent with the setup in \cite{Ref_Chapter5}.

\vspace{-8pt}
\subsection{Performance Evaluation in Grid-Connected Mode}

In this scenario, the microgrid maintains a connection to the main grid via the PCC. The load profile (Fig. \ref{fig:grid_load}) is designed with sharp ramps to stress the system dynamics and force specific GFL units towards their saturation boundaries.

First, the proposed strategy incorporating the active constraint handling mechanism (Activation Function $\bm{A}^{ac}$) is evaluated. As illustrated in Fig. \ref{fig:grid_withac_P}, the output powers of all DGs adjust smoothly to track the load changes. A critical event occurs between $t=2$s and $t=4$s, where the load demand drives GFL units 2 and 4 to their upper active power limit of 0.5 MW. 
Crucially, despite the saturation, no oscillations are observed. This stability is attributed to the activation function, which detects the boundary condition and logically isolates these nodes from the consensus update loop, as per Eq. \eqref{eq:activation}. Consequently, the standardized power increments for the unsaturated DGs (Fig. \ref{fig:grid_withac_Pd}) continue to converge within a narrow error band of 0.09. This indicates that the remaining resources (GFM units and PCC) automatically and proportionally compensate for the power deficit left by the saturated GFLs, strictly adhering to the tertiary economic ratios.
Simultaneously, the multi-scale attention mechanism successfully filters out the FDI attacks initiated at $t=2$s, preventing any divergence in the consensus variables. Reactive power sharing (Figs. \ref{fig:grid_withac_Q} and \ref{fig:grid_withac_Qd}) exhibits similar coordinated behavior, with a slightly larger error band (0.37) due to the localized voltage support, yet maintaining global stability.

For comparative analysis, the simulation is repeated using a traditional consensus controller that lacks the activation function logic. The results, depicted in Fig. \ref{fig:grid_withoutac}, reveal significant performance degradation. Without the mechanism to handle saturation, the controller continues to integrate the error for the saturated GFL units (DGs 2 and 4). This leads to the "integrator wind-up" phenomenon, causing the standardized increments to diverge significantly, with deviations reaching 0.41—a 355\% increase compared to the proposed method (Fig. \ref{fig:grid_withoutac_Pd}).
Physically, this divergence forces the GFM units and the PCC to undergo excessive transient oscillations to restore balance. As shown in Fig. \ref{fig:grid_withoutac_P}, the overshoot during the recovery phase (post $t=4$s) is approximately 3.8 times larger than that of the proposed strategy. These oscillations not only stress the physical components but also threaten the stability of the entire grid connection. This comparison provides compelling evidence that the proposed active constraint handling mechanism is essential for the safe and stable operation of hybrid microgrids.

\vspace{-8pt}
\subsection{Robustness Verification in Islanded Mode}

In the islanded scenario, the PCC is disconnected, forcing the microgrid to operate autonomously. This imposes stricter requirements on the GFM units to maintain voltage and frequency while balancing power. To test the system's limits, a more aggressive load fluctuation profile with steeper ramps is applied (Fig. \ref{fig:island_load}).

The dynamic response of the proposed strategy is shown in Fig. \ref{fig:island_withac}. Despite the severity of the load disturbances and the absence of the main grid, the system maintains robust performance. The standardized active power increments (Fig. \ref{fig:island_withac_Pd}) converge within an error band of 0.15. This confirms that the GFM units effectively act as the slack bus references, dynamically sharing the load while respecting the operational limits of the GFL units.
A key observation in islanded mode is the relaxation of strict reactive power consensus. As shown in Fig. \ref{fig:island_withac_Qd}, a larger deviation in reactive increments is observed between GFL and GFM units. This is a deliberate design feature: in islanded operation, GFM units prioritize voltage stability over precise reactive power sharing with GFLs. However, the GFM units themselves maintain tight synchronization (error $< 0.036$), ensuring high-quality voltage support.

Conversely, the traditional strategy fails to maintain effective coordination in the islanded mode, as evidenced by Fig. \ref{fig:island_withoutac}. The divergence in standardized increments widens uncontrollably to 0.68 for active power and 0.73 for reactive power. A critical instability is observed around $t=5$s (Fig. \ref{fig:island_withoutac_P}), where the charging power of the GFM units surges abruptly. This instability arises because the controller attempts to force saturated GFL units to track an infeasible consensus value, causing the GFM units to overcompensate wildly. In a real-world scenario, such a surge would likely trigger protection relays and cause a blackout. These results confirm that the proposed hierarchical strategy significantly enhances the resilience and stability of hybrid microgrids, preventing system collapse in vulnerable islanded operations.

\section{Conclusion}
\label{sec:concl}

This paper presents a resilient hierarchical power control strategy for hybrid GFL/GFM microgrids, effectively bridging the gap between economic optimization and dynamic regulation while addressing physical and cyber constraints. By introducing the standardized power increment as a unified interface, the proposed framework ensures that real-time load fluctuations are shared strictly according to tertiary economic dispatch ratios, resolving timescale conflicts. Crucially, the integration of a dynamic activation mechanism with projection operators actively manages GFL saturation, preventing integrator wind-up and preserving system stability even when individual units reach their physical limits. Furthermore, the robust secondary control design, fortified with multi-scale attention and predictive compensation, guarantees Uniformly Ultimately Bounded stability against unbounded FDI attacks and packet losses. Simulation results on an IEEE 33-bus system confirm that the strategy significantly enhances coordination accuracy and operational resilience in both grid-connected and islanded modes. Future work will explore fully decentralized tertiary optimization to further eliminate single points of failure and enhance scalability.







\bibliographystyle{IEEEtran} 
\bibliography{bibtex/bib/GFML} 






\end{document}